\documentclass[lettersize,journal]{IEEEtran}
\usepackage{amsmath,amsfonts,amssymb,stmaryrd}
\usepackage{xfrac}
\usepackage{array}
\usepackage[caption=false,font=footnotesize,labelfont=sf,textfont=sf]{subfig}
\usepackage{textcomp}
\usepackage{stfloats}
\usepackage{url}
\usepackage{verbatim}
\usepackage{graphicx}
\usepackage{authblk}
\def\BibTeX{{\rm B\kern-.05em{\sc i\kern-.025em b}\kern-.08em
    T\kern-.1667em\lower.7ex\hbox{E}\kern-.125emX}}
\usepackage{balance}
\usepackage{pdfpages}

\usepackage{color,soul}
\usepackage{threeparttable}
\usepackage{xcolor}
\usepackage[hidelinks]{hyperref}
\usepackage{adjustbox}
\usepackage{float}
\definecolor{Paired-1}{RGB}{31,120,180}
\definecolor{Paired-2}{RGB}{166,206,227}
\definecolor{Paired-3}{RGB}{51,160,44}
\definecolor{Paired-4}{RGB}{178,223,138}
\definecolor{Paired-5}{RGB}{227,26,28}
\definecolor{Paired-6}{RGB}{251,154,153}
\definecolor{Paired-7}{RGB}{255,127,0}
\definecolor{Paired-8}{RGB}{253,191,111}
\definecolor{Paired-9}{RGB}{106,61,154}
\definecolor{Paired-10}{RGB}{202,178,214}
\definecolor{Paired-11}{RGB}{177,89,40}
\definecolor{Paired-12}{RGB}{255,255,153}
\definecolor{dgreen}{RGB}{3,125,60}
\definecolor{dblue}{RGB}{47,82,143}
\definecolor{bgreen}{RGB}{0,128,128}
\definecolor{bgred}{RGB}{237,29,33}
\usepackage{bm}
\usepackage{dsfont}
\usepackage{numprint}

\usepackage{adjustbox}
\usepackage{booktabs}
\usepackage{multirow}

\usepackage{cite}
\usepackage{algpseudocode}
\usepackage{caption}
\usepackage{balance}
\usepackage[utf8]{inputenc}

\usepackage{graphicx}
\usepackage{textcomp}
\usepackage{pgfplots, pgfplotstable}
\usepackage{tikz}
\usetikzlibrary{spy} 
\usetikzlibrary{matrix, positioning, patterns, shapes, arrows}
\usetikzlibrary{positioning}
\usepackage{tikzscale}
\usepgfplotslibrary{statistics}
\usepackage{bchart}
\pgfplotsset{compat=newest}
\usepgfplotslibrary{groupplots}
\usepgfplotslibrary{colorbrewer}
\usepackage[linesnumbered,ruled,vlined]{algorithm2e}

\SetCommentSty{mycommfont}
\usepackage{lipsum}

\usepackage{microtype}

\renewenvironment{thebibliography}[1]{\section{References}\small\list
 {[\arabic{enumi}]}{\settowidth\labelwidth{[#1]}\leftmargin\labelwidth
 \advance\leftmargin\labelsep
 \usecounter{enumi}}
 \def\newblock{\hskip .11em plus .33em minus .07em}
 \sloppy\clubpenalty4000\widowpenalty4000
 \sfcode`\.=1000\relax}

\usepackage{etoolbox}
\patchcmd{\thebibliography}{\leftmargin\labelwidth}
    {\itemsep -1.4pt \leftmargin\labelwidth}{}{}{}

\pgfplotsset{every x tick label/.append style={font=\scriptsize}}
\pgfplotsset{every y tick label/.append style={font=\scriptsize}}

\makeatletter
\newcommand\thefontsize[1]{{#1 The current font size is: \f@size pt\par}}
\makeatother

\makeatletter
\def\bstctlcite{\@ifnextchar[{\@bstctlcite}{\@bstctlcite[@auxout]}}
\def\@bstctlcite[#1]#2{\@bsphack
  \@for\@citeb:=#2\do{%
    \edef\@citeb{\expandafter\@firstofone\@citeb}%
    \if@filesw\immediate\write\csname #1\endcsname{\string\citation{\@citeb}}\fi}%
  \@esphack}
\makeatother 

\newcommand{\sign}{\text{sgn}}

\def\BibTeX{{\rm B\kern-.05em{\sc i\kern-.025em b}\kern-.08em
    T\kern-.1667em\lower.7ex\hbox{E}\kern-.125emX}}

\graphicspath{{}}

\newcommand{\conftitle}{\footnotesize SUBMITTED TO IEEE TRANSACTIONS ON CIRCUITS AND SYSTEMS-I IN JULY 2023 – REVISED IN AUGUST 2023}
\usepackage{fancyhdr}

\fancypagestyle{pageStyleOneAndOther}{%
    \fancyhf{}
    \fancyhead[R]{\conftitle}
}

\begin{document}
\title{Pipelined Architecture for Soft-decision Iterative Projection Aggregation Decoding for RM~Codes}



\author{\IEEEauthorblockN{Marzieh Hashemipour-Nazari\IEEEauthorrefmark{1}, Yuqing Ren\IEEEauthorrefmark{2}, Kees Goossens\IEEEauthorrefmark{1}, and Alexios Balatsoukas-Stimming\IEEEauthorrefmark{1}\\}
              \IEEEauthorblockA{\IEEEauthorrefmark{1}Electronic Systems, Eindhoven University of Technology, The Netherlands\\}
              \IEEEauthorblockA{\IEEEauthorrefmark{2}Telecommunications Circuits Laboratory, \'Ecole Polytechnique F\'ed\'erale de Lausanne, Switzerland}
              }
\maketitle

\thispagestyle{pageStyleOneAndOther}
\pagestyle{pageStyleOneAndOther}

\begin{abstract}
The recently proposed recursive projection-aggregation (RPA) decoding algorithm for Reed-Muller codes has received significant attention as it provides near-ML decoding performance at reasonable complexity for short codes. However, its complicated structure makes it unsuitable for hardware implementation. Iterative projection-aggregation (IPA) decoding is a modified version of RPA decoding that simplifies the hardware implementation. In this work, we present a flexible hardware architecture for the IPA decoder that can be configured from fully-sequential to fully-parallel, thus making it suitable for a wide range of applications with different constraints and resource budgets. Our simulation and implementation results show that the IPA decoder has $41\%$ lower area consumption, $44\%$ lower latency, four times higher throughput, { but currently seven times higher power consumption} for a code with block length of $128$ and information length of $29$ compared to a state-of-the-art polar successive cancellation list (SCL) decoder with comparable decoding performance.
\end{abstract}


\section{Introduction}
\label{intro}
\IEEEPARstart{F}{uture} communications systems will need to enable ultra-reliable low-latency communications (URLLC) and machine-type communications (MTC)\cite{Mahmood2020}.
Low latency generally implies the use of very short packets. Moreover, in some MTC systems, such as Internet of Things (IoT) applications, there is not enough data to create large packets because sensors typically only transmit a small amount of data infrequently~\cite{Durisi2016,Chen2018}. 
{ Low-density parity check (LDPC)~\cite{gallager1963low} and turbo~\cite{berrou1996near} codes are highly regarded due to their ability to achieve significant coding gains in the moderate blocklength regime, while maintaining linear decoding complexity.
However, conventional asymptotic methods used to construct LDPC and turbo-like codes, such as extrinsic information transfer (EXIT) charts, often have difficulties to generate short codes with good performance.
As a result, achieving high reliability with short packets becomes challenging, as conventional error-correcting schemes typically require moderate to large blocklengths to be effective.}

{An alternative approach for short packets is to utilize polar codes, which are capacity-achieving codes with low encoding and decoding complexity for binary-input memoryless symmetric (BMS) channels  under successive cancellation (SC) decoding~\cite{Arikan2009}.
However, to make {polar} codes effective for short blocklengths, certain modifications are necessary.
Typically, this means employing the SC list (SCL) decoder with a very large list size, combined with a CRC code.
Consequently, this results in a reduced effective information rate for the code and increased decoding complexity and latency.
These challenges have increased attention towards codes and decoding algorithms specifically designed for short-length packets~\cite{Coskun2019,Tonnellier2021}, aiming to enhance communication performance and achieve low latency.}


 
{Reed-Muller (RM) codes are a class of linear block error-correcting codes that are closely related to polar codes. They were first discovered and introduced by Reed~\cite{Reed1954} and Muller~\cite{Muller1954}.}
Reed's decoder is based on majority voting and can correct a number of errors up to half of the code's minimum distance. Several additional decoding methods were developed to improve the decoding performance~\cite{Dumer2004,Sakkour2005,Dumer2006a,Dumer2006,saptharishi2017}. 
More recently, there has been a renewed interest in RM codes as they can achieve the Shannon capacity on any BMS channel~\cite{Costello2007,Abbe2015,Kudekar2017,Abbe2019,Sberlo2020} and they were shown to outperform polar codes under maximum likelihood (ML) decoding for short codes~\cite{Arkan2009,Mondelli2014}.





As ML decoding is generally intractable, the authors of \cite{Ye2020} introduced a more practical near-ML decoding method for RM codes called recursive projection-aggregation (RPA) decoding. 
The RPA algorithm exploits the recursive structure of RM codes by projecting a received codeword with a length of $n$ into $n-1$ shorter codewords and decoding the projected codewords recursively until codewords belonging to a first-order RM code are reached, which can be decoded efficiently using the fast Hadamard transform. 
The number of recursive calls depends on the order of the employed RM code. The complexity of RPA decoding scales as $n^r$, where $r$ is the order and $n$ is the length of the RM code. However, decoding low-order RM codes (i.e., second and third-order RM codes) with a short length is still practical with RPA decoding, making it particularly interesting for URLLC and MTC applications.
Nevertheless, the complexity and recursive structure of RPA decoding are still major challenges for its efficient hardware implementation.

Some modified versions of the RPA algorithm have been proposed to reduce its algorithmic complexity. Simplified RPA~\cite{Ye2020} is a variant of RPA deploying two-dimensional sub-spaces for the projection step, which reduces the total number of projections. The authors of~\cite{Lian2020} proposed a collapsed projection-aggregation (CPA) decoding algorithm, which merges multiple recursion levels into a single step and has lower complexity. The results in~\cite{Lian2020} show that the CPA algorithm achieves a similar error-correcting performance to the RPA algorithm for RM codes with $r=3$ and $n=128$. 
To further reduce complexity,~\cite{Huang2022} and \cite{li2022optimization} proposed different ways to exploit correlations between projection to prune CPA.   
Although both simplified RPA and CPA reduce the overall algorithmic complexity, they make the projection and aggregation steps more involved as they employ more complex operations.
Sparse RPA (SRPA)~\cite{Fathollahi2021} is another modification of the RPA decoder that consists of multiple sparse RPA decoders. Each sparse RPA decoder uses only a random subset of projections. The work of \cite{JiaJie2021} has lowered the average computational complexity of RPA by taking advantage of syndrome-based early stopping techniques along with a scheduling scheme. 

Even though all aforementioned algorithmic complexity-reduction methods for RPA are promising, there are still challenges in their hardware implementation due to their recursive and/or complex structure. The iterative projection-aggregation (IPA)~\cite{Hashemipour-Nazari2021} algorithm transforms the recursive structure of the RPA decoder into an iterative structure, making it more straightforward for hardware implementation. The work of \cite{Hashemipour-Nazari2021} includes a preliminary fully-parallel hardware implementation of the IPA algorithm, but only for the special case of hard-decision decoding. 


\subsubsection*{Contributions} The main contributions of this paper are:
\begin{itemize}
\item We first design a flexible processing unit for soft-decision  IPA decoding that can perform one level of projection, first-order decoding, and a part of the aggregation step. The proposed processing unit is configurable at runtime for performing different projections and their corresponding aggregations. Moreover, we propose hardware-friendly architectures for the projection and aggregation steps that reduce the required hardware resources and simplify the data flow. 	
   
\item We design a flexible pipelined architecture based on the proposed processing units for the {{IPA}} algorithm that can be configured to be from fully-sequential to fully-parallel. As a result, the proposed architecture is very flexible in trading latency and throughput for area,
 such that it is applicable to a wide range of URLLC and MTC systems with different requirements and constraints.
{Additionally, to achieve high throughput, we design the controlling path of the architecture to support pipelining. This enables efficient data processing and optimal utilization of available resources.}


\item We compare the error-correcting performance of short RM codes under IPA decoding with similar 5G polar codes under SCL decoding~\cite{Tal2015,Balatsoukas-Stimming2015}. Moreover, we compare our proposed architecture for the IPA decoder and a state-of-the-art implementation of the SCL decoding~\cite{Ren2022} for the same blocklength, rate, and error-correcting performance, and { we show that our IPA decoder is superior with respect to the area and latency in the short blocklength and low-rate regime. However, it currently has a higher power consumption compared to a state-of-the-art SCL decoder}.
\end{itemize}

\subsubsection*{Outline} The remainder of this paper is organized as follows.
In Section~\ref{backgnd}, we review the background of RM codes as well as the RPA and IPA decoding algorithms. In Section ~\ref{sec:sipa_2nd}, we provide a detailed description of our proposed architecture for the second-order IPA decoder by explaining the structure of the processing unit, voting circuit, register array, and control unit. In Section~\ref{sec:gen}, we explain how our basic second-order decoder architecture can be generalized to decode RM codes of any order. In Section~\ref{sec:result}, we discuss the simulation and implementation results. We first compare the error-correcting performance of the IPA decoder to the baseline RPA decoder and then present multiple simulations to justify certain parameter choices for the hardware implementation in Section~\ref{sec:sim_res}. In addition, we compare the hardware implementation results of the IPA decoder for different RM codes with hard-decision (HD) IPA~\cite{Hashemipour-Nazari2021}, and a state-of-the-art SCL decoder~\cite{Ren2022} for polar codes in Section~\ref{sec:impl_res}.  
Finally, Section \ref{sec:conclusion} concludes this paper.

\section{Background}
\label{backgnd}
\subsubsection*{Notation} In this manuscript, lowercase and uppercase boldface letters denote vectors and matrices, respectively. In addition, vectors of log-likelihood ratios (LLR) are denoted by the boldface uppercase and non-italic letter $\mathbf{{L}}$. The symbols  $ \mathbf{y}^i$ and $y(j)$ represent the $i$-th projected vector and the $j$-th coordinate of the vector $ \mathbf{y}$, respectively. 

\subsection{Reed-Muller codes}
Reed-Muller (RM) codes are linear block codes denoted by RM$(m,r)$ with rate $R=\dfrac{k}{n}$, where $n=2^m$ indicates the code length, $r$ is the order, and ${k=\sum _{i=0}^r \binom{m}{i}}$ is the dimension of the code.
There are several ways to define RM codes~\cite{MacWilliams,Abbe2021} including a recursive approach, which is called Plotkin construction. 
In the  Plotkin $(\mathbf{u},\mathbf{u}+\mathbf{v})$ construction of RM codes, $\mathbf{u}$ and $\mathbf{u}+\mathbf{v}$ are two subvectors of length $2^{m-1}$  of a codeword $\mathbf{c} = (\mathbf{u},\mathbf{u}+\mathbf{v}) \in \text{RM}(m,r)$, where $\mathbf{u} \in \text{RM}(m-1,r)$ and $\mathbf{v} \in \text{RM}(m-1,r-1)$.
Therefore, the generator matrix $\mathbf{G}_{(m, r)}\in \mathbb{F}^{k\times n} $ for an $\text{RM}(m,r)$ code is recursively defined based on the Plotkin construction as: 
\begin{equation}\label{eq:GMtxRM}
\mathbf{G}_{(m, r)}= \begin{bmatrix}
  \mathbf{G}_{(m-1, r)} & \mathbf{G}_{(m-1, r)}\\ 
  \mathbf{0} & \mathbf{G}_{(m-1, r-1)}
\end{bmatrix},
\quad
\mathbf{G}_{(1, 1)}{=} \begin{bmatrix}
  1 & 1\\ 
  0 & 1
\end{bmatrix}.
\end{equation}
In addition, the minimum Hamming distance of the RM$(m,r)$ code is $d = 2^{m-r}$.
	
\subsection{First-order RM codes and decoder}
\label{sec:fht}
RM$(m,1)$ represents the first-order RM code with length $n=2^m$, dimension $k=m+1$, and minimum Hamming distance $d=2^{m-1}$.
The most popular optimal decoding method for first-order RM codes is the decoding algorithm based on the fast Hadamard transform (FHT)~\cite{Green66,Beext86}.
The FHT-based decoding operates in the following four steps: 
\begin{enumerate}
\item Calculate the FHT of the received vector $\mathbf{y}$, represented by its LLR values $\mathbf{L}$ as:
	\begin{equation} \label{eq:fhtrans}
		\mathrm{\omega} = \mathbf{H}_{2^m}\mathbf{L},
	\end{equation}
where the Hadamard matrix $\mathbf{H}_{2^m}$ is defined as:
\begin{equation} \label{eq:fhtmatrix}
\mathbf{H}_{2^m} = \begin{bmatrix}
  \mathbf{H}_{2^{m-1}} & \mathbf{H}_{2^{m-1}}\\ 
  \mathbf{H}_{2^{m-1}} & -\mathbf{H}_{2^{m-1}} 
\end{bmatrix} ,
\: \: \mathbf{H}_2 = \begin{bmatrix}
1 & 1\\
1 & -1
\end{bmatrix}.
\end{equation}
\item Find the index  $\beta$ of vector $\mathrm{\omega}$ such that: 
\begin{align}
	\beta = \arg\max _{i\in \{0,\hdots,n-1\}} |\omega (i)|
\end{align}
\item Calculate the index $\alpha$ of the closest codeword $\hat{\mathbf{y}}\in \text{RM}(m,1)$ to the received vector as:
		\begin{equation} \label{eq:fhtind}
		\alpha = n\lambda+\beta\text{ where } \lambda = \begin{cases}0, & \omega(\beta) \geq 0, \\ 1, & \omega(\beta)<0.\end{cases}
	\end{equation}
\item The decoded codeword $\hat{\mathbf{y}}\in \text{RM}(m,1)$ is then given by:
\begin{equation}
	\hat{\mathbf{y}}=\operatorname{de2bi}\left(\alpha\right)\mathbf{G}_{(m,1)},
\end{equation}
where $\operatorname{de2bi}\left(\alpha\right)$ gives the right-MSB binary representation of the index $\alpha$.
\end{enumerate}


\subsection{RPA decoding}
\label{sec:rpa}
As mentioned in Section~\ref{intro}, a wide variety of decoding algorithms have been proposed for RM codes, some of which make use of the recursive structure and the large automorphism group of RM codes to propose projection-based methods. 
One of the algorithms taking advantage of the recursive structure of RM codes is RPA decoding. As shown in Fig.~\ref{fig:RPA}, the RPA method uses three steps, namely, projection, recursive decoding, and aggregation to decode a noisy received vector $\mathbf{y}$ from a transmitted RM codeword $\mathbf{c}$.

In addition, the RPA algorithm has two flavors: 1) hard-decision decoding where $\mathbf{y}$ is a binary vector, mostly used for binary symmetric channels (BSCs), and 2) soft-decision decoding where $\mathbf{y}$ is a vector of LLRs, which can be used for more general communication channels like additive white Gaussian noise (AWGN) channels. 
{The general structure of the RPA algorithm is the same for both hard- and soft-decision decoding, but the projection and aggregation steps are different.} In this paper, we focus on soft-decision decoding.
RPA decoding can be described by the following steps:


\begin{figure}[t]
  \centering
  \centerline{\includegraphics[width=0.49\textwidth]{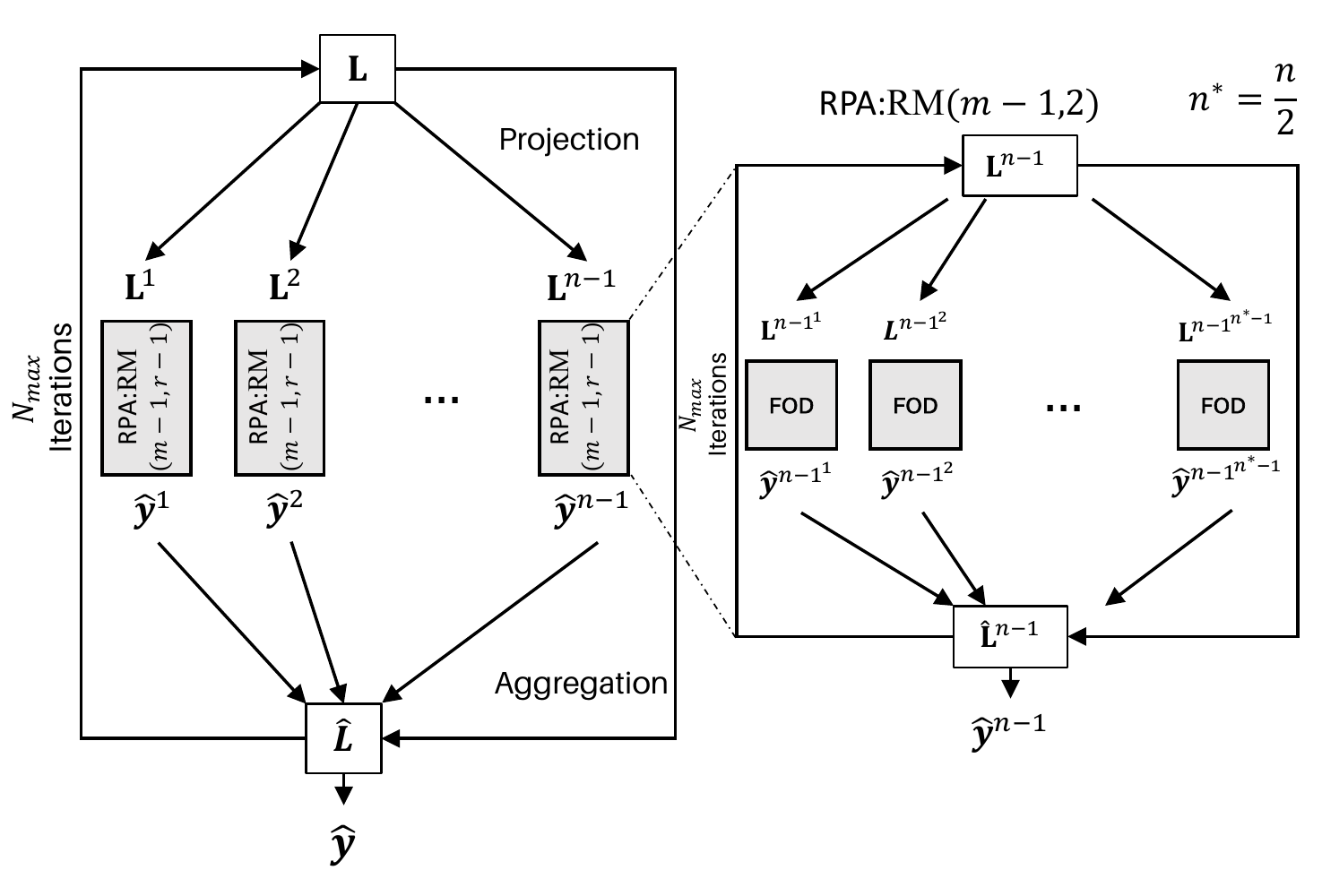}}
  \caption{\small RPA decoding for third-order RM codes based on~\cite{Ye2020}.}
  \label{fig:RPA}
\end{figure} 

\subsubsection{Projection}
\label{projstep}

In this step, $\mathbf{L}$ is transformed into $n-1$ distinct vectors $\mathbf{L}^1, \mathbf{L}^2,...,\mathbf{L}^{n-1}$ of length $\dfrac{n}{2}$.
The projection rule is:
\noindent
\begin{equation} \label{eq:tanh}
\resizebox{.9\hsize}{!}
{$\mathbf{L}^i(j) = 2 
\tanh ^{{-}1}\left(
\tanh \left(\frac{\mathbf{L}\left(j^a\right)}{2}\right)
\tanh \left(\frac{\mathbf{L}\left(j^b\right)}{2}\right) 
\right),$}
\end{equation}
where $i \in \{0,\hdots,n-1\}$ is the projection number, $j\in \{0,\hdots,\frac{n}{2}-1\}$, and $j^a$ and $j^b$ are the coordinates of the original vector $\mathbf{L}$ that are used to create $\mathbf{L}^i(j)$. The set of these coordinates for the $i$-th projection is given by:
\begin{equation}
\label{eq:xorcoor}
  \begin{array}{c}
    \lbrace (j^a,j^b)\vert j^a=j^b \oplus i; \forall j^b \in \{0,\hdots,n-1\} \rbrace.
  \end{array}
\end{equation}
We note that the above set contains $\sfrac{n}{2}$ pairs of elements for which $(j^a,j^b) = (j^b,j^a)$. To avoid repetition, for each such pair, we remove $(j^b,j^a)$ from the set.
Equation \eqref{eq:tanh} is often approximated by the so-called min-sum approximation~\cite{Fossorier1999}, which is defined as:
\begin{align} \label{eq:MinSum}
\begin{split}
\mathbf{L}^i(j) = {}&\min\left\lbrace|\mathbf{L}(j^a)|,|\mathbf{L}(j^b)|\right\rbrace\sign\big(\mathbf{L}(j^a)\big)\sign\big(\mathbf{L}(j^b)\big). 
\end{split}
\end{align}

\subsubsection{Recursive decoding}
In this step, each projected vector from the previous step is recursively decoded with RPA decoding for RM$(m-1,r-1)$ until first-order codes are reached, for which the FHT-based first-order decoder (FOD) explained in Section~\ref{sec:fht} is applied.



\subsubsection{Aggregation}
\label{sec:agg}
In this step, a per-coordinate average is taken from all the decoded codewords obtained from the recursive decoding step. Then, the hard-decoded binary vector $\mathbf{\hat{y}}$  from the obtained LLR vector $\hat{\mathbf{L}}$ is considered as an estimation for the transmitted codeword $\mathbf{c}$.
This step is effectively the reverse of the projection step, so similarly to the projection step, it requires first finding the origins of each coordinate $\hat{y}^i(j)$, $j \in \{1,\hdots,n/2\}$, of the decoded codeword $\mathbf{\hat{y}}^i$, $i \in \{1,\hdots,n-1\}$. 
The {\ttfamily{RevReorder}} function given in Algorithm~\ref{alg:rev_reorder} finds the index pair $\mathbf{U}^i(j):=(j^a,j^b)$ for each coordinate $\hat{y}^i(j)$, that was originally created by the $i$-th projection on the coordinates $j^a$ and $j^b$ of the input vector $\mathbf{L}$. Then, the average of the LLR values that were involved in creating each pair of coordinates ${\hat{y}}(j^a)$ and ${\hat{y}}(j^b)$ is calculated as follows:
\begin{align}
\label{eq:agg}
{\hat{\mathbf{L}}}(j^a) & =\frac{1}{n-1}\sum_{i=1}^{n-1}\left(1-2\hat{y}^i(j) \right)\mathbf{L}(j^b), \\
{\hat{\mathbf{L}}}(j^b) & =\frac{1}{n-1}\sum_{i=1}^{n-1}\left(1-2\hat{y}^i(j) \right)\mathbf{L}(j^a).
\end{align}  
Furthermore, as shown in Fig.~\ref{fig:RPA}, several iterations of the aforementioned steps are performed at every recursion level until either there are no changes in the decoded codeword or a pre-defined maximum number of iterations $N_{\max}$ is reached, which the authors of \cite{Ye2020} set to$\lceil {m/2}\rceil$.

\begin{algorithm}[t]
\SetAlgoLined
\textbf{Input: } $i, n $ \\
\textbf{Output: }\resizebox{.8\hsize}{!}{$\mathbf{U}=\{\mathbf{u}(j)\mid\mathbf{u}(j)=(j^a,j^b); j=0,...,\frac{n}{2}-1\}$} \\

\eIf{$i < \dfrac{n}{2}$}{
$\mathbf{U}\left(0 \; \text{to} \;\dfrac{n}{4}{-}1\right)\gets ${\ttfamily{RevReorder}}$\left(i,\frac{n}{2}\right)$\\
$\mathbf{U}\left(\dfrac{n}{4} \; \text{to} \;\dfrac{n}{2}{-}1\right)\gets ${\ttfamily{RevReorder}}$\left(i,\frac{n}{2}\right)+\dfrac{n}{2}$
}{
\For{$j{=}1$ : $\dfrac{n}{2}{-}1$}{
$\mathbf{U}(j) \gets \left(j,j\oplus i\right)$ \\
}
$\mathbf{U}(0) \gets (0,i)$
}

return $\mathbf{U}$
\caption{{\ttfamily{RevReorder}}}
\label{alg:rev_reorder}
\end{algorithm}

\subsection{IPA decoding and implementation}
\begin{figure*}[!t]
\centering
\subfloat[]{\includegraphics[width=0.54\textwidth]{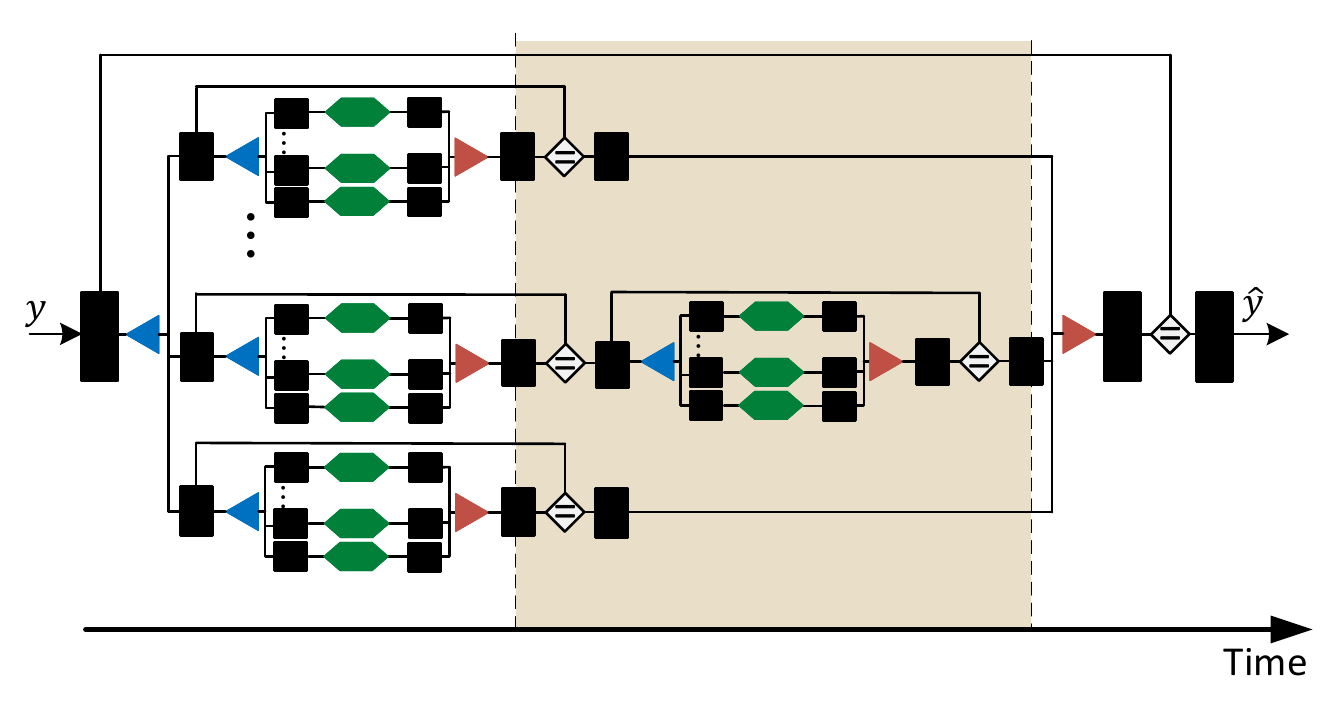}%
\label{fig_df_rpa}}
\hfil
\subfloat[]{\includegraphics[width=0.46\textwidth]{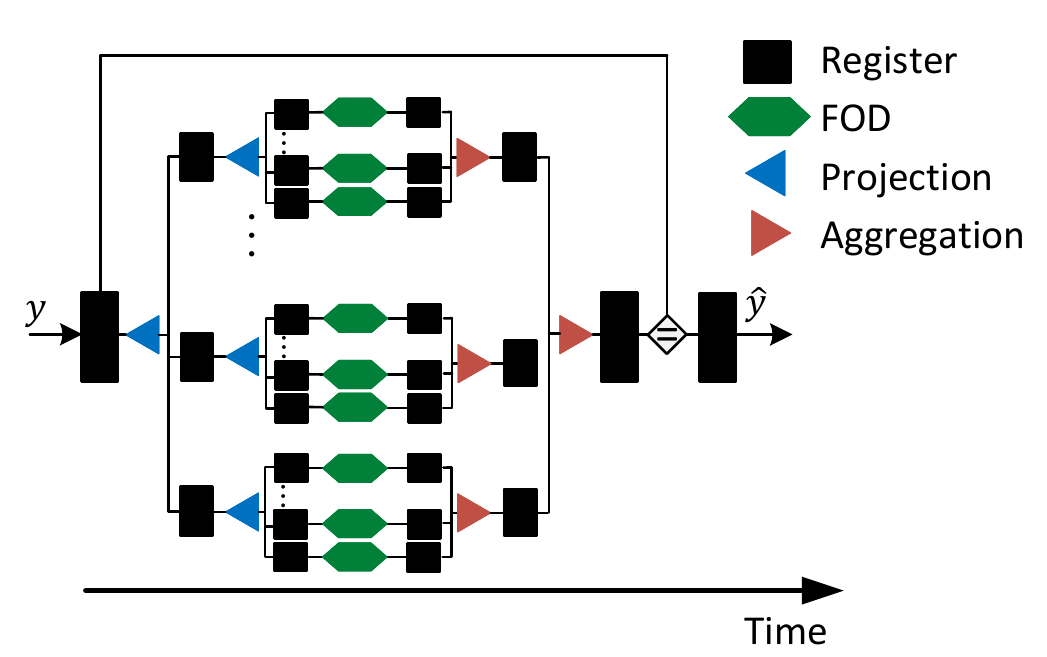}%
\label{fig_df_ipa}}
\caption{Data-flow of one iteration of RPA (a) and IPA (b) decoding of an arbitrary codeword from $RM(m,3)$ code.}
\label{fig:Dflow}
\end{figure*}

RPA decoding performs up to $N_{\max}$ iterations at each level of the recursion, which increases the complexity and significantly complicates the RPA decoding structure.
Fig.~\ref{fig_df_rpa} shows the data flow of one iteration of the RPA algorithm to decode a codeword from an $\text{RM}(m,3)$ code. 
In this example, two levels of projection are shown with blue triangles until the first-order codes are reached. Then, the FODs, represented with green hexagons, decode the first-order RM codes. Similar to the projection step, there are two levels of aggregation represented with brown triangles. However, as shown in the highlighted region, the second projected vector requires an extra iteration on its recursive call to the RPA of $RM(m-1,2)$ after the first level of aggregation. This makes the other parallel branches stall. 
Furthermore, handling multiple iterations at each recursion level requires complicated control circuitry and memory structure when implemented in hardware.

The work of \cite{Hashemipour-Nazari2021} showed that it is possible to remove the iterations at internal levels of recursion with minimal degradation in error-correcting performance.
This change transforms the recursive structure of RPA decoding into an iterative one that we call IPA, shown in Fig.~\ref{fig_df_ipa}.
This simplification of the RPA algorithm made it feasible to have a fully-parallel hardware implementation~\cite{Hashemipour-Nazari2021}. 
{
\subsubsection*{Computational complexity}
The number of calls to the FOD function is commonly used as a measure to estimate the computational complexity of RPA decoding~\cite{Lian2020,Huang2022,li2022optimization,Fathollahi2021,JiaJie2021}. 
In the worst-case scenario, where the maximum number of iterations is performed for every recursive call, the total number of FOD calls can be determined as:
\begin{equation} \label{eq:Complexity}
 \Theta_{\left(m,r,N_{\max}\right)}=N_{\max}^{r-1}\prod_{i=1}^{r-1}\left(2^{m-i-1}-1\right).
\end{equation}
However, with the simplification in IPA, the number of FODs will be decreased by $N_{\max}^{r-2}$ times as there is no additional iteration for $r-2$ levels of recursion.
}
\subsubsection*{Implementation}
{ The proposed architecture in~\cite{Hashemipour-Nazari2021} is comprised of three main components: \emph{projection}, \emph{FOD}, and \emph{aggregation} for HD decoding of the received vector $\mathbf{y}\in \text{RM}(m,r)$.
The \emph{projection} component performs {$r-1$} levels of projection.
Each projection level is equipped with parallel projection units. These units are comprised of a crossbar to combine the corresponding coordinates for the desired projection of the input vector $\mathbf{y}$. Additionally, each unit includes an XOR circuits to apply the projection rule for HD decoding.
The \emph{FOD} component provides first-order decoding for all first-order codewords,  obtained in the innermost level of projection, in parallel.
Each \emph{FOD} consists of the hardware implementation of each step in the decoding method explained in Section~\ref{sec:fht}.  
The \emph{aggregation} component provides $r{-}1$ levels of aggregation. Each level includes the required parallel crossbars to expand the corresponding coordinates of the desired decoded codeword from RM$(m-j,r-j)$ in order to estimate the codeword belonging to RM$(m-j+1,r-j+1)$, where $j=\{1,\ldots,r-1\}$ represents the current level of aggregation.}

Although the proposed architecture has very low latency, the resource utilization is extremely high due to its fully-parallel structure.
{Moreover, the decoder only supports HD decoding, which accepts binary vectors as input rather than LLR values. The projection and aggregation rules for HD decoding, which are explained in detail in \cite{Ye2020} and \cite{Hashemipour-Nazari2021}, differ significantly from those of a soft-decision decoder.
}

{ In Section~\ref{sec:sipa_2nd}, we present a flexible hardware implementation of the soft-decision IPA decoding for general binary-input memoryless channels. Our design can be easily configured based on the specific requirements of the application, offering a range of options from fully sequential to fully parallel configurations including partial-parallel configuration.}

\begin{algorithm}[t]
\SetAlgoLined
\textbf{Input: } $\mathbf{L}, m, N_{\max} $ \\
\textbf{Output: }Codeword $\hat{\mathbf{y}}$\\
 $n \gets 2^m$ \\
\For {$j=0:N_{\max}$}{
\For{$i{=}1$ : $n{-}1$}{
$\mathbf{L}^i \gets${\ttfamily{Projection}}$(\mathbf{L},m,i)$ \\
$\hat{\mathbf{y}}^i\leftarrow$ {\ttfamily{FOD}}$\left(\mathbf{L}^i\right)$ \tcp{First-order decoder}
$\mathbf{L}_{\text{agg}}^i \gets$ {\small{\ttfamily{PreAggregation}}}$(\mathbf{L},\hat{\mathbf{y}}^i,i,m)$ \\
}
 $\mathbf{L} \gets $ {\ttfamily{Voting}}$\left(\mathbf{L}_{\text{agg}}^1,...,\mathbf{L}_{\text{agg}}^{n-1}\right)$\\
}
\For {$z=0 : n-1$}{
 $ \hat{\mathbf{y}}\gets$ $1 - 1 \left[\mathbf{L}(z))\geq 0 \right]$ \tcp{Hard-decision}
}
return  $\hat{\mathbf{y}}$\\
\caption{{\ttfamily{IPA}} decoding for RM$(m,2)$ codes}
\label{alg:2nd_IPA}
\end{algorithm}


{
\subsection{Polar code and SCL decoder}
\label{sec:polar}
Polar codes~\cite{Arikan2009} were ratified as the channel coding scheme of 5G enhanced mobile broadband (eMBB).
However, even with highly specialized node-based SCL decoders capable of parallel bit decoding, meeting the stringent demands of high-reliability and low-latency in 5G (and beyond) is still a challenge for existing polar decoders.
Recently, the authors of~\cite{Ren2022} proposed the first generalized node-based SCL decoding algorithm and presented a corresponding hardware implementation.
By extending a generalized node called the \emph{sequence repetition} (SR) node\cite{Zheng21} to SCL decoding, this state-of-the-art polar decoder in~\cite{Ren2022} achieves increased decoding parallelism and more efficient utilization of computation units. This enhancement offers a competitive solution for 5G polar codes. We compare our proposed architecture against this state-of-the-art polar decoder in Section~\ref{sec:result}.
}

\section{Second-order soft-input IPA decoder architecture}
\label{sec:sipa_2nd}
\begin{figure*}[!t]
\centering
\includegraphics[width=0.75\textwidth]{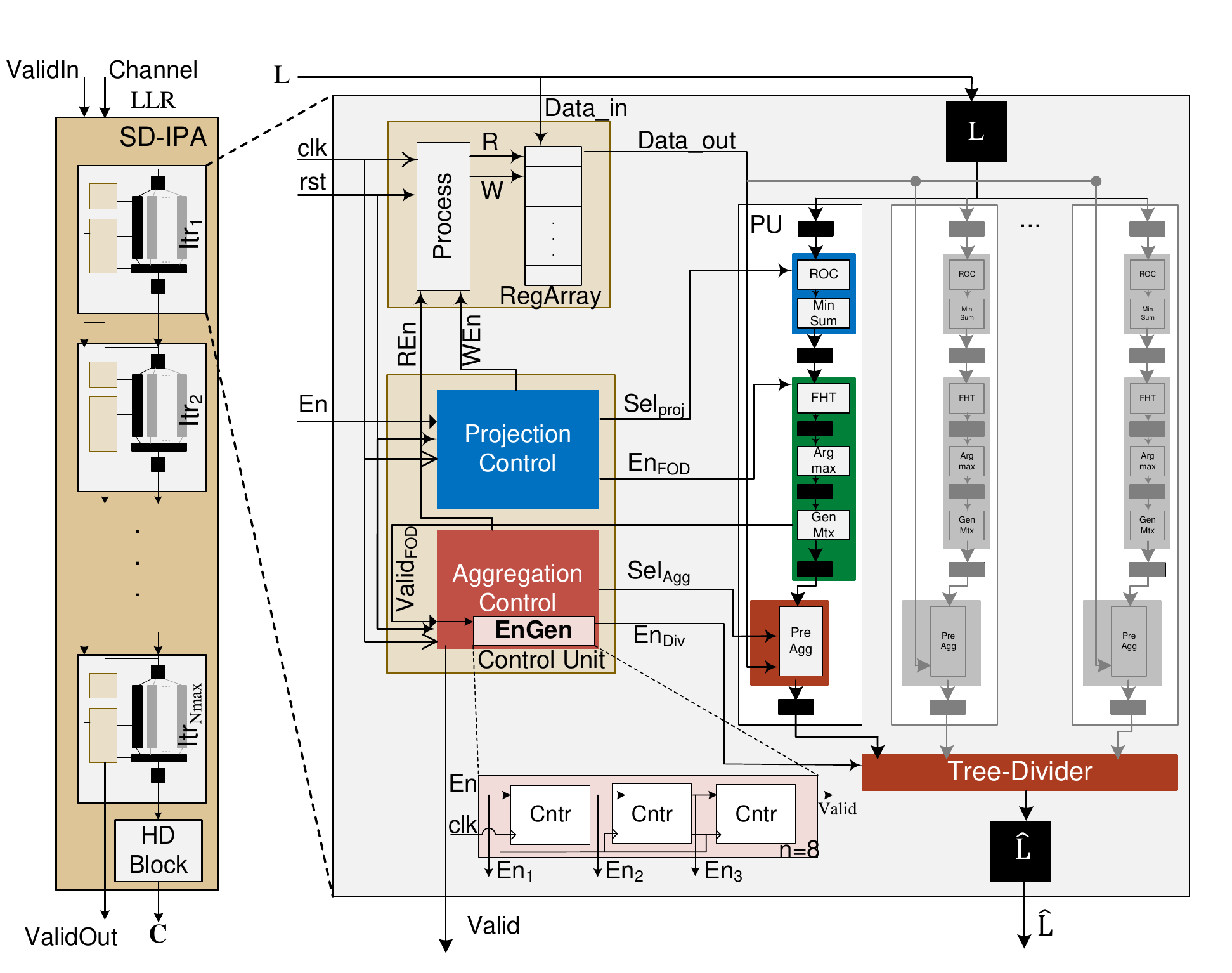}
\caption{Overview of our proposed second-order SIPA decoder architecture. }
\label{fig:SIPA_2nd}
\end{figure*}
Since first-order RM codes can be decoded optimally using the FHT~\cite{Beext86}, the lowest order for which RPA decoding is meaningful is $r=2$.
Therefore, we first describe a base architecture for IPA decoding for second-order RM codes and we explain how it can be extended to decode RM codes of higher order in Section~\ref{sec:gen}. 

We implement {soft-decision} IPA decoding as described in Algorithm~\ref{alg:2nd_IPA} with a pipelined architecture shown in Fig.~\ref{fig:SIPA_2nd} that has dedicated hardware blocks for each iteration of the outer for loop in lines 4-11 of Algorithm~\ref{alg:2nd_IPA}.
The inputs of this architecture are a vector of channel LLRs and the \textit{ValidIn} signal, which is high when a new vector of channel LLRs becomes available. The outputs are the decoded codeword and the \textit{ValidOut} signal, which is high when the output is valid. Furthermore, our architecture includes processing units (PUs) that implement the projection step, first-order decoding, and a part of the aggregation step, which we call pre-aggregation.
A hardware-friendly pipelined tree divider is also placed after the PU to complete the aggregation step of the IPA decoding. In the following subsections, we explain these components in detail.

\subsection{Processing unit (PU)}
The for loop in lines 5-9 of Algorithm~\ref{alg:2nd_IPA} can be fully parallelized.
Therefore, we design a PU which is pipelined and has appropriate hardware components to perform the {\ttfamily{Projection}}, {\ttfamily{FOD}}, and {\ttfamily{PreAggregation}} functions for any value of the loop variable $i\in\{0,\hdots,n-1\}$, where $i=0$ corresponds to a dummy all-zeros vector that simplifies the implementation and that we explain in Section~\ref{sec:div}. Thus, our second-order IPA decoder can be from fully-sequential with $P=1$ PUs to fully-parallel with $P=2^m$ PUs. 

\subsubsection{Projection component}
\begin{figure*}[!t]
\centering
\includegraphics[width=0.7\textwidth]{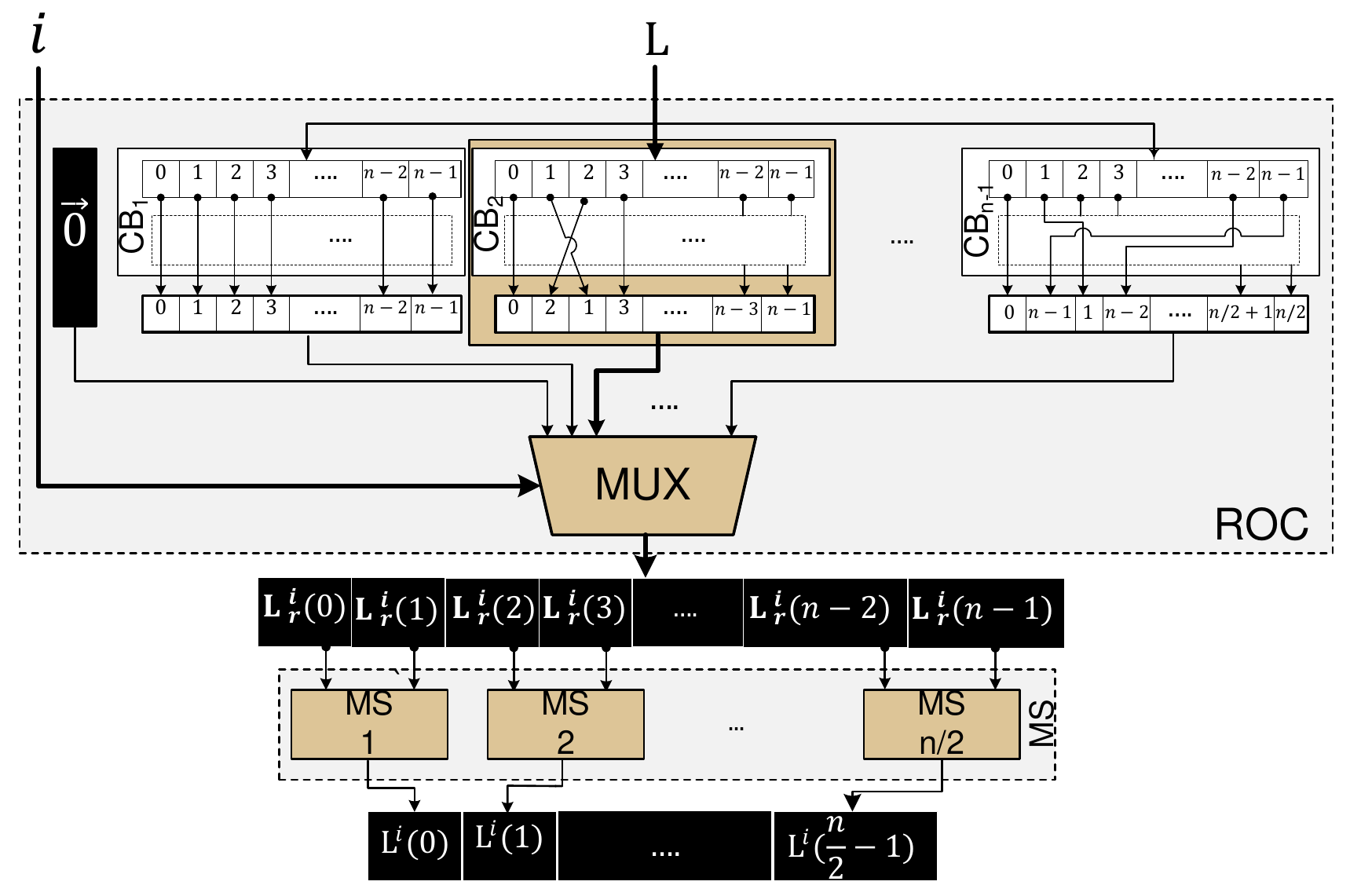}
\caption{Architecture of the \textit{Projection} component including the ROC and \textit{MS} sub-components.}
\label{fig:proj}
\end{figure*}

This component includes two sub-components: Reordering (\textit{ROC}) and min-sum (\textit{MS}) shown in Fig.~\ref{fig:proj}. 
The ROC contains $n-1$ crossbars (\textit{CB}) and a multiplexer that selects the crossbar for the corresponding projection.
Each crossbar $i$ is built with function {\ttfamily{Reorder}}$\left(\mathbf{L}, i, m\right)$ represented in Algorithm~\ref{alg:reorder} that finds the relevant coordinates of the input vector $\mathbf{L}$, taking the projection number $i$ into account.
Then, it reorders $\mathbf{L}$ to put those coordinates in consecutive pairs. Consequently, the output vector $\mathbf{L}^i_r$ is a reordered version of the input vector $\mathbf{L}$ according to the projection number $i$. 
{As shown in Fig.~\ref{fig:proj}, the multiplexer selects the reordered vector $\mathbf{L}^i_r$  corresponding to the current projection $i$ determined with its \textit{selector} input.}
For example, in Fig.~\ref{fig:proj}, the highlighted crossbar shows that the second projection is currently performed.
\begin{algorithm}[t]
\SetAlgoLined
\textbf{Input: } $\mathbf{L}, i, n $ \\
\textbf{Output: }$\mathbf{L}^i_r$\\
\eIf{$i < \dfrac{n}{2}$}{
	$\mathbf{L}^i_r(0 \; \text{to} \;\dfrac{n}{2}{-}1)\gets ${\ttfamily{Reorder}}$( \mathbf{L}(0 \; \text{to} \;\dfrac{n}{2}{-}1),i,m{-}1)$ \\
	$\mathbf{L}^i_r(\dfrac{n}{2} \; \text{to} \;n-1) \gets ${\ttfamily{Reorder}}$(\mathbf{L}(\frac{n}{2}\; \text{to} \;n{-}1),i,m{-}1)$
}
{
	\For {$j{=}1$ : $\dfrac{n}{2}{-}1$}{
		$\mathbf{L}^i_r(2j) \gets \mathbf{L}(j)$ \\
		$\mathbf{L}^i_r(2j+1) \gets \mathbf{L}(j\oplus i)$
	}
	$\mathbf{L}^i_r(0) \gets \mathbf{L}(0)$\\
	$\mathbf{L}^i_r(1) \gets \mathbf{L}(i)$
}
return $\mathbf{L}^i_r$
\caption{{\ttfamily{Reorder}}}
\label{alg:reorder}

\end{algorithm}

The number of crossbars inside the \textit{ROC} block depends on the number of available PUs. When $P$ PUs are instantiated, the \textit{ROC} block placed in the $j$-th PU, $j\in \{0,\hdots,P-1\}$, includes the crossbars $i$ such that $i \mod P =j$. Therefore, for one PU, the \textit{ROC} block contains all the crossbars, but for $P>1$, each PU has a distinct circuit for its \textit{ROC} sub-component. In addition, the \textit{MS} component shown in Fig.~\ref{fig:proj} consists of $\sfrac{n}{2}$ blocks performing \eqref{eq:MinSum} on every pair of coordinates $\left(\mathbf{L}^i_r(2j),\mathbf{L}^i_r(2j+1)\right), j\in \left\{0,\hdots,\sfrac{n}{2}-1\right\},$ of the reordered vector $\mathbf{L}_r^i$. 


\begin{figure}[!t]
\centering
\includegraphics[width=0.4\textwidth]{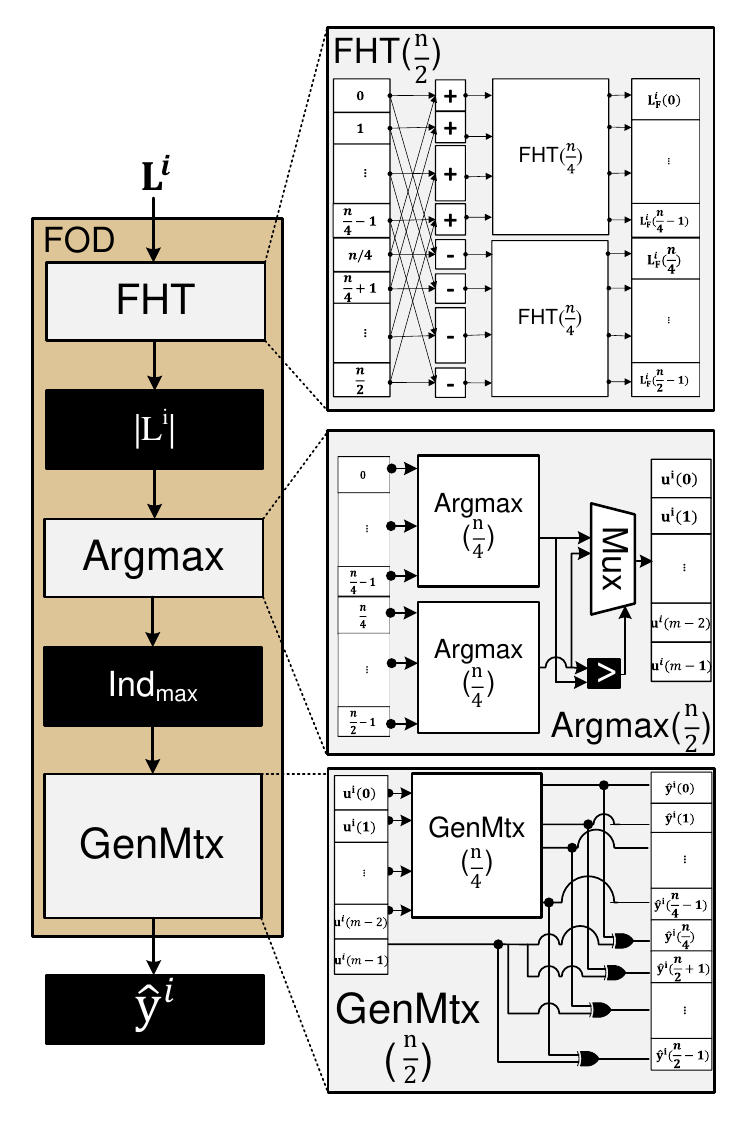}
\caption{Hardware implementation of FOD based on \cite{Hashemipour-Nazari2021} }
\label{fig:SFOD}
\end{figure}
\subsubsection{FOD component}
We use the architecture proposed in \cite{Hashemipour-Nazari2021} modified for soft-input FOD, explained in Section~\ref{sec:fht}, to decode the vector $\mathbf{L}^i$ corresponding to a first-order codeword obtained from the \textit{Projection} component. As Fig.~\ref{fig:SFOD} demonstrates, the \textit{FOD} component is pipelined and contains three modules: 1) the \textit{FHT} module, which computes the fast Hadamard transform of $\mathbf{L}^i$, 2) the \textit{Argmax} module to find the index of the maximum value of the output of the \textit{FHT} module, and 3) the \textit{GenMtx} module, which implements an encoder for an RM$(m-1,1)$ code. The output of the \textit{FOD} component is the decoded codeword $\hat{\mathbf{y}}^i$ corresponding to the projected vector $\mathbf{L}^i$.

\subsubsection{PreAggregation component}
\begin{figure*}[t]
\centering
\includegraphics[width=0.9\textwidth]{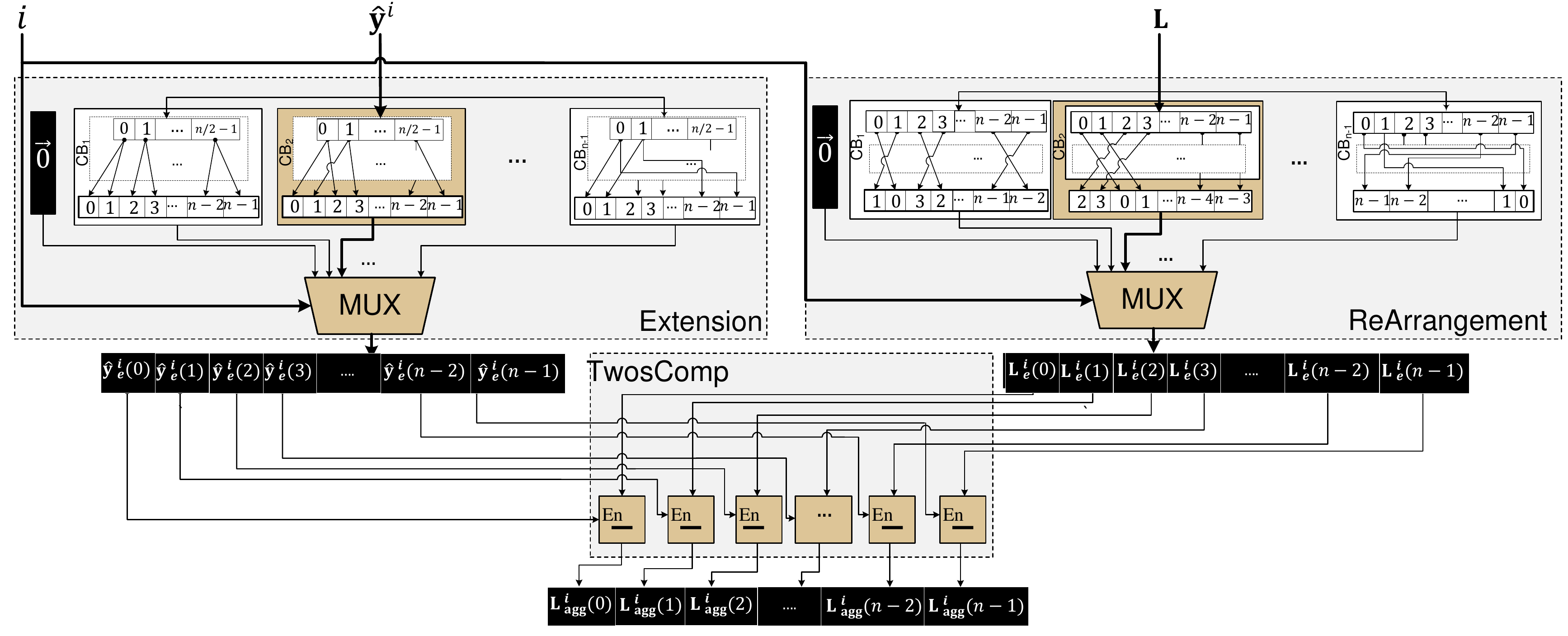}
\caption{Hardware implementation of the {\ttfamily{PreAggregation}} function.}
\label{fig:preagg}
\end{figure*}

\begin{algorithm}[t]
\SetAlgoLined
\textbf{Input: } $\mathbf{L},\boldsymbol{\hat{y}}^1,\boldsymbol{\hat{y}}^2,\hdots,\boldsymbol{\hat{y}}^{n-1}, m $  \\
\textbf{Output: }$\hat{\mathbf{L}}$ \\
 $n \gets 2^m$ \\
\For{$i{=}1$ : $n{-}1$}{
	\tcp{{\ttfamily{PreAggregation}}$(\mathbf{L},\hat{\boldsymbol{y}}^i,i,m)$}
	$\mathbf{U}^i \gets$\ttfamily{RevReorder}$(i,m)$  \\
		\For {$j=0$ : $\frac{n}{2}-1$}{
			$(j^a,j^b) \gets \mathbf{U}^i(j)$ \\
			 $\mathbf{L}_{\text{agg}}^i(j^a) \gets \left(1-2\hat{y}^i(j) \right) \mathbf{L}(j^b))$ \\
			 $\mathbf{L}_{\text{agg}}^i(j^b) \gets \left(1-2\hat{y}^i(j) \right) \mathbf{L}(j^a))$ \\
		}
}
\tcp{{\ttfamily{Voting}}}
\For {$z=0$ : $n-1$}  {
	$\hat{\mathbf{L}}(z) \gets \frac{\sum_{i=1}^{n-1}\mathbf{L}_{\text{agg}}^i(z)}{n-1}$
	}

\caption{{\ttfamily{Aggregation}}}
\label{alg:Agg_sd}
\end{algorithm}
We have rewritten the aggregation step explained in Section~\ref{sec:agg} in Algorithm~\ref{alg:Agg_sd}. The for loop on lines 4-11 of Algorithm~\ref{alg:Agg_sd} can be fully parallelized. The {\ttfamily{PreAggregation}}$(\mathbf{L},\hat{\mathbf{y}}^i,i,m)$ function in Algorithm~\ref{alg:2nd_IPA} implements exactly this for loop.
The {\ttfamily{PreAggregation}} function first calls {\ttfamily{RevReorder}} function corresponding to the $i$-th loop variable for finding the pairs of indices $(j^a,j^b)$ of the input vector $\mathbf{L}$ from which the coordinates $\mathbf{L}^i(j), j\in\{0,\hdots,\sfrac{n}{2}\}$ were originally created. Then, it applies the aggregation rule in ~\eqref{eq:agg}, except the final averaging step.
The hardware implementation of the {\ttfamily{PreAggregation}} function, shown in Fig.~\ref{fig:preagg}, has three sub-components: 
\begin{itemize}
\item \textit{Extension} module, which extends the length-$\sfrac{n}{2}$ decoded binary codeword $\mathbf{\hat{y}}^i$ to the length-$n$ vector $\mathbf{\hat{y}}^i_{\text{e}}$. The extension rule is determined by the {\ttfamily{RevReorder}} function, which copies each coordinate ${\hat{y}}^i(j), j\in\{0,\hdots,\sfrac{n}{2}\}$, to their corresponding coordinates of $\mathbf{\hat{y}}^i_{\text{e}}$ indexed by $j^a$ and $j^b$.

\item \textit{ReArrangement} module, which flips the values of each pair of coordinates $(\mathbf{L}(j^a),\mathbf{L}(j^b))$. The output vector with flipped LLRs is $\mathbf{L}_\text{e}^{i}$. 
Similarly to the \textit{Extension} module, each crossbar $i$ in the \textit{ReArrangement} module finds the pair $(j^a,j^b)$ based on the {\ttfamily{RevReorder}} function.   

\item \textit{TwosComp} module, which finds the two's complement of each coordinate $j,j\in\{0,\hdots,n-1\}$, of $\mathbf{L}_\text{e}^{i}$ if ${\hat{{y}}}^i_{\text{e}}(j)=1$, as stated in lines 8-9 of Algorithm~\ref{alg:Agg_sd}. 
\end{itemize}
In addition, similarly to the \textit{ROC} module, both the \textit{Extension} and the \textit{ReArrangement} modules contain a multiplexer to select the extension network or crossbar corresponding to the current decoded projected vector $\mathbf{\hat{y}}^i$. Moreover, the number of the extension networks and crossbars inside the \textit{Extension} and  \textit{ReArrangement} modules is adjusted based on the number of instantiated PUs, similarly to the \textit{ROC} module.

\subsection{Tree divider}
\label{sec:div}
The second for loop in the {\ttfamily{Aggregation}} function, shown in lines 12-14 of Algorithm~\ref{alg:Agg_sd},
is the {\ttfamily{Voting}} function mentioned in Algorithm~\ref{alg:2nd_IPA}. 
More specifically, it takes the average value of the pre-aggregated vectors $\mathbf{L}_{\text{agg}}^i,i\in \{1,\hdots,n-1\}$, which are the outputs of PUs. 
As shown in line 11 of Algorithm~\ref{alg:Agg_sd}, the  {\ttfamily{Voting}} function needs an $(n-1)$-input adder to add up all $\mathbf{L}_{\text{agg}}^i$, $i=1,\hdots,n-1$. 
In addition, it requires a divider to find the average value out of the accumulated LLRs. Implementing such adder and divider in hardware is relatively expensive, especially when $n$ gets large. As a result, we propose a tree divider structure to implement the  {\ttfamily{Voting}} function.

\begin{figure}[t]
\centering
\includegraphics[width=0.37\textwidth]{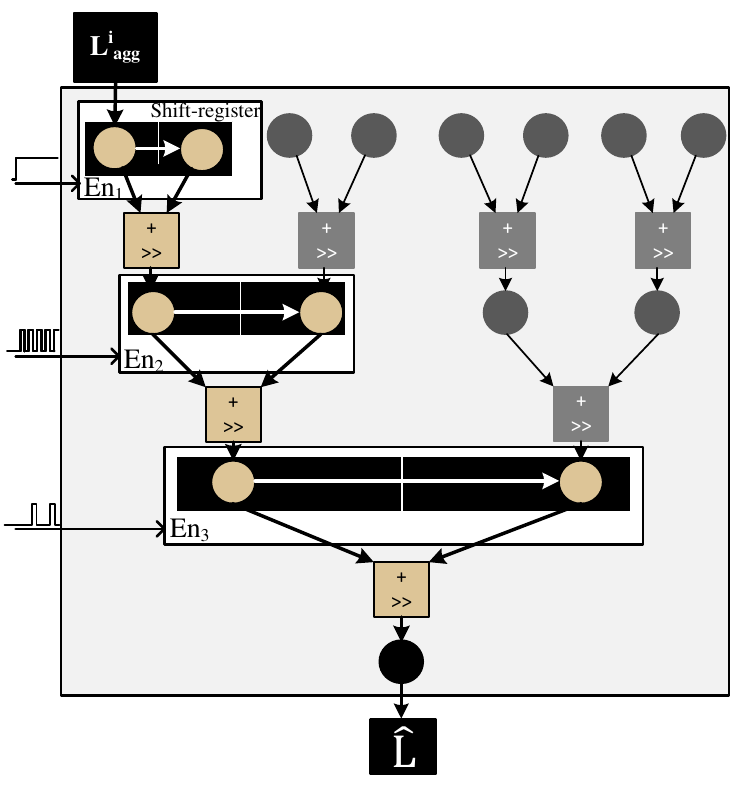}
\caption{Hardware implementation of the {\ttfamily{Voting}} function with a tree-like divider for an  example of $n=8$.}
\label{fig:divider}
\end{figure}

The structure of the divider follows from the fact that the average value of two sets of $\sfrac{n}{2}$ numbers is equal to the sum of the average values of each set divided by two:
\begin{equation} \label{eq:adder}
\bar{x}=\frac{1}{n} \sum_{i=1}^{n}x_i   =\frac{ \frac{1}{\frac{n}{2}}\sum_{i=1}^{\frac{n}{2}}x_i +  \frac{1}{\frac{n}{2}} \sum_{i=\frac{n}{2}+1}^{n}x_i }{2}.
\end{equation}
As a result, if $n=2^m$, we end up with a tree structure represented in Fig.~\ref{fig:divider}. In this structure, we have two-input adders with one bit extension and shifting elements that perform division by two, simply shifting the dividend one bit to the right. 
However, there are $n-1$ pre-aggregated vectors instead of $n$ required vectors for the tree divider. Consequently, we add a dummy all zero vector to the flow of our proposed decoder. This approximation affects the output of {\ttfamily{Voting}} function, but the effect is negligible, as we will show in Section~\ref{sec:sim_res}.
The $\mathbf{0}$ vector is generated with $i=0$ in \textit{Projection} and \textit{PreAggregation} units as it is shown in Fig.~\ref{fig:proj} and Fig.~\ref{fig:preagg}.

In the case of a fully-sequential implementation with $P=1$  PUs, only one new divider input is available per clock cycle. Hence, we implemented the divider, depicted in Fig.~\ref{fig:divider}, with $\log(n)$ shift registers, two-input adders, and shifting elements. Each shift register contains two registers for two LLRs, data and enable input ports, as well as one data output port. The shift register shifts its internal array one position to the right and writes its current data input to the location of the shifted value. 
For the shift register in the first level $l=0$,  the data input is the output of the PU, and its enable is the valid output port of the PU. The data input for the shift register in level $l>0$ is the output of the shift register in level $l-1$, and its enable input is high when the shift register at level $l-l$ has written two new data inputs. This signal is generated with \textit{EnGen} module in the \textit{Control} unit.  

However, in a partially-parallel design with $P>1$ PUs, more than one input is ready at every clock cycle. Taking the tree divider's structure into account, we add a condition for choosing the number of PUs to keep the divider and pipeline stages as simple as possible. This constraint limits $P=2^p, p\in\{0,\hdots,m\}$, so that we replace the shift registers with standard registers for the first $p$ levels of the tree divider. Hence, we have $2^{p-l}$ two-input standard registers at each level $l$, $l\in \{0, \hdots, m-p-1\}$. These registers are enabled with the valid output of the \textit{PreAggregation} unit delayed by $l-1$ clock cycles. Furthermore, we keep the sequential divider operating with shift registers for the last $m-p$ levels of the divider. The shift register placed in the first level of the sequential part which is the $(m-p)$-th level is enabled with the valid signal of \textit{PreAgg} unit delayed by  $p$ clock cycles. The enable signals for the rest of the shift registers are generated by the \textit{Control} unit.

\subsection{Register array}
\begin{figure*}[!t]
\centering
\includegraphics[width=\textwidth]{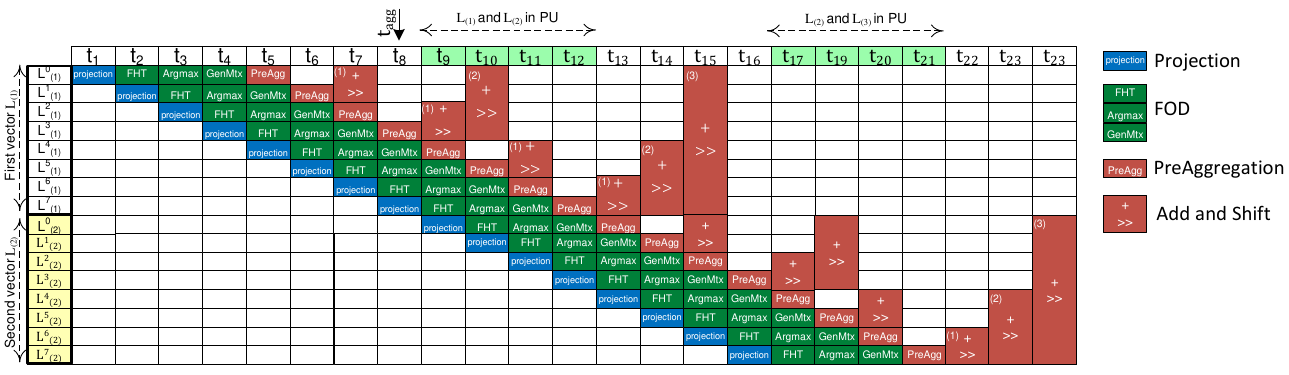}
\caption{Pipeline stages for the fully-sequential second-order IPA decoder for an example of $n=8$. }
\label{fig:stages}
\end{figure*}

Fig.~\ref{fig:stages} demonstrates the pipeline stages for a  fully-sequential implementation of IPA, i.e., $P=1$ , for RM$(3,2)$. In this example, for simplicity each stage is assumed to require one clock cycle. However, in the hardware implementation, the length of each stage can be multiple clock cycles to balance the critical path depending on the input blocklength.
As shown in the highlighted time slots in Fig.~\ref{fig:stages}, the \textit{Projection} component loads the new received vector while the \textit{PreAggregation} component is still busy with the previously received vector. Therefore, we need { an array of registers} to store the channel LLRs for each input vector $\mathbf{L}$ during the decoding process.
The depth of this {array} is calculated by:
\begin{equation} \label{eq:mem}
D_{\text{RegArr}} = \bigg\lceil{\frac{t_{\text{agg}}}{n/P}}\bigg\rceil+1,
\end{equation}
where $t_{\text{agg}}$ is the number of clock cycles required for the pipeline stages before reaching the \textit{PreAggregation} component, $P$ is the number of PUs, and $n$ is the block length.
Hence, the more PUs we instantiate, the more {registers} we need. In the case of a fully-sequential implementation with $P=1$, only two registers are required to store the channel LLRs of two input vectors as $n > t_{\text{agg}}$ in most practical cases. On the other hand, in a fully-parallel implementation with $P=n$ PUs, we have $D_{\text{RegArr}} = {t_{\text{agg}}}+1$. As a result, this register array is generally relatively small.
{The register array is managed using a small \textit{write} counter, which keeps track of the register that needs to be updated. Additionally, a counter is utilized to determine the register that needs to be read when the \textit{PreAggregation} component begins processing a new vector. The \textit{Control} unit generates the necessary signals to facilitate the updating and reading of registers within the array.}

\subsection{Control unit}
The \textit{Control} unit generates all the controlling signals required for the pipeline structure, the PUs, the tree divider, and the memory. According to the pipeline table represented in Fig.~\ref{fig:stages} for an implementation that is not fully parallel, it is not possible to decode one codeword every clock cycle, so the \textit{read} and \textit{write} counters in the memory should be controlled carefully. Besides, the selectors for the multiplexers in the \textit{Projection} and \textit{PreAggregation} components of the PU do not have the same value, e.g., when the \textit{PreAggregation} starts processing the first decoded projected codeword at $t_5$, the \textit{Projection} starts the $5$-th projection in the example shown in Fig.~\ref{fig:stages}. Moreover, as mentioned earlier, the data input $\mathbf{L}$ is not the same for the \textit{Projection} and \textit{PreAggregation} units in one PU as shown in the highlighted part from time instances $t_9$ to $t_{12}$ and from $t_{17}$ to $t_{21}$ in Fig.~\ref{fig:stages}. Therefore, the \textit{Control} unit runs two state machines simultaneously to generate the controlling signals for the projection and aggregation steps.

In addition, the \textit{EnGen} component of the \textit{Control} unit generates the enable signals for the registers used in the tree divider. It consists of $m-p$ cascaded counters for implementing the IPA decoder for the RM$(m,2)$ code with $2^p$ PUs.
Each counter makes its output high for one clock cycle when it counts two high levels of its enable signal. The enable signal of the first counter is the valid output of the PU delayed by $p$ clock cycles. Furthermore, the remaining counters in the consecutive levels are enabled with the previous level's output. 
    
\subsection{Iteration}
After the aggregation step in the RPA algorithm, a comparison is made to check whether the output of the current iteration converges to its input or not. If the condition is satisfied, the algorithm stops before performing all $N_{\max}$ iterations. 
However, since very few iterations are generally required, we removed the early stopping condition in our proposed architecture to simplify the pipeline flow and to have a constant throughput.

For a decoder with $N_{\max}$ iterations, we cascade $N_{\max}$ copies of the single-iteration hardware, as shown in Fig.~\ref{fig:SIPA_2nd}, such that the \textit{ValidIn} input port of $i$-th iteration is the \textit{ValidOut} of $(i-1)$-th iteration. As Fig.~\ref{fig:SIPA_2nd} shows, there is a hard-decision maker module at the end of the flow. This module considers the most-significant bit of the estimated LLRs of the $N_{\max}$-th iteration as the binary decoded codeword.     

\subsection{Throughput and latency}
As shown in Fig.~\ref{fig:stages}, a new codeword with blocklength $n$ can be processed every $n$ clock cycles in a fully-sequential IPA decoder. However, in the partially-parallel IPA decoder with $P$ PUs, a new codeword can be inserted into the pipeline every $\frac{n}{P}$ clock cycles. In general, the throughput of the proposed second-order IPA decoder is:
\begin{equation} \label{eq:thr}
{\text{Thr}_{\text{Mbps}}} = \frac{P \times \text{Frequency}}{n}\times n = P\times \text{Frequency},
\end{equation}
where the frequency is given in MHz.

As all steps of the IPA decoder are pipelined, the latency of one iteration of the second-order IPA decoder is:
\begin{equation} \label{eq:latency}
{{t}_{(m,2)}} = \left(t_{\text{proj}}+t_{\text{FOD}}+t_{\text{PreAgg}}\right)+\left(\frac{2^m}{P}-1\right)+t_{\text{Divider}}+t_{\text{IO}},
\end{equation}
where $t_{\text{proj}}$, $t_{\text{FOD}}$, and $t_{\text{PreAgg}}$ are the delays of the \textit{Projection}, \textit{FOD}, and \textit{PreAggregation} components, respectively, measured in clock cycles. In addition, $t_{\text{Divider}}=m$ and $t_{\text{IO}}=2$ corresponding to two input and output registers. In our design, we consider $t_{\text{proj}}={t_\text{PreAgg}}=1$, and $t_{\text{FOD}}=3$ or $t_{\text{FOD}}=4$, depending on the input blocklength. The total latency of the second-order IPA decoder with $N_{\max}$ iterations is ${{t}_{(m,2)}}\times N_{\max}$.
 
\section{Proposed architecture for soft-input IPA decoder for RM codes with $r>2$}
\label{sec:gen}
\begin{figure}[t]
\centering
\includegraphics[width=0.525\textwidth]{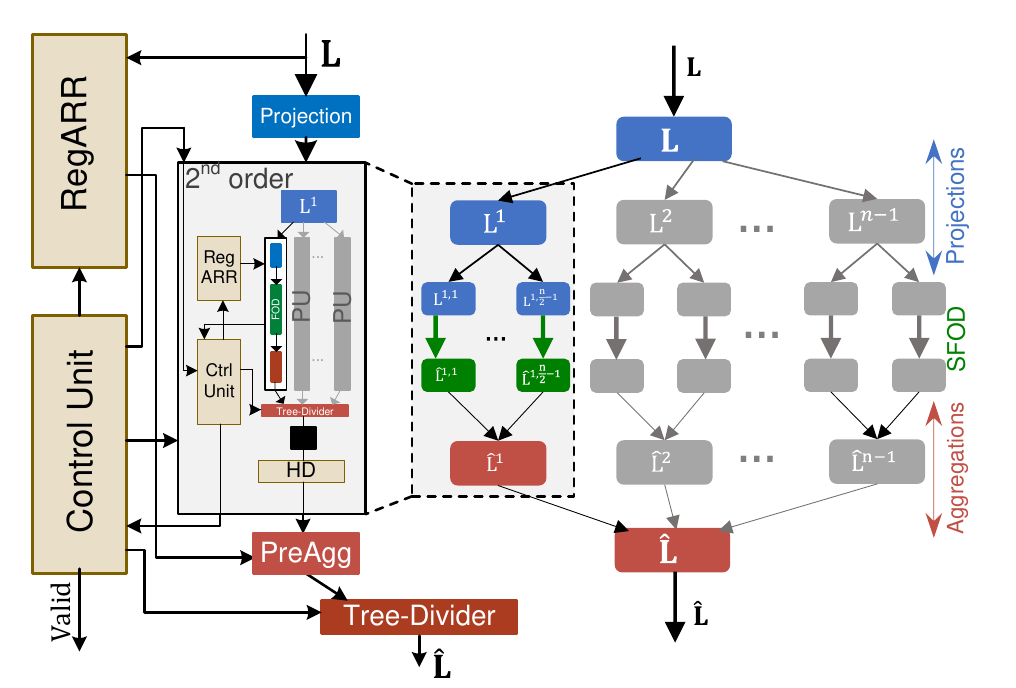}
\caption{Overview of third-order IPA decoder architecture for  one iteration. }
\label{fig:SIPA_3rd}
\end{figure}

IPA decodes RM$(m,r)$ codes with $r>2$ by producing $2^m-1$ projected codewords from RM$(m-1,r-1)$, as discussed in Section~\ref{sec:rpa}. Therefore, we can generalize the architecture shown in Fig.~\ref{fig:SIPA_2nd} to decode RM codes with $r>2$ by adding \textit{Projection} and \textit{PreAggregation} components along with the tree divider for any level $r>2$. Additionally, a dedicated memory with different blocklength and depth is required for each level of $r>2$. Finally, we add a \textit{Control} unit for each level of $r>2$ to generate the control signals required for the corresponding $r$. Fig.~\ref{fig:SIPA_3rd} shows an overview of the proposed architecture for one iteration of the third-order IPA decoder.   

We keep the dummy all-zero vector inserted in the base second-order IPA decoder, and the projection, first-order decoding, and aggregation steps are performed on this dummy vector. As a result, some clock cycles are lost, but it simplifies the divider and \textit{Control} unit as mentioned in Section~\ref{sec:sipa_2nd}.
However, for the levels $r>2$, we do not use the dummy all-zero vector because it results in wasting significant number of clock cycles depending on the structure of the base second-order decoder. Therefore, to keep the tree divider and the whole structure simple, we only freeze the entire decoder for one clock cycle at each level of $r>2$ and we insert zeros in the frozen cycle to the tree divider directly.

Similarly to the second-order IPA decoder, the proposed $r$-order IPA decoder is able to be adjusted from fully-sequential to fully-parallel. A fully-sequential architecture is obtained by instantiating one PU in the level $r=2$ and a fully-parallel implementation is obtained by instantiating all $\left(\prod_{i=0}^{r-3} 2^{m-i}-1\right)\times 2^{m-r+2}$ required PUs for the level $r=2$. Any other number of available PUs, following the constraint mentioned in Section~\ref{sec:div}, results in partially-parallel architectures. It is worth mentioning that with $P>2^{m-r+2}$, i.e., with more than the required PUs for a fully-parallel base second-order IPA, multiple second-order IPA decoders will be instantiated in the design.

Similar to the second-order IPA decoder, the throughput of the higher-order IPA decoder is calculated as:
\begin{equation} \label{eq:thr3}
{\text{Thr}_{\text{bps}}} = \frac{P \times \text{Frequency}}{\left(\prod_{i=0}^{r-3} 2^{m-i}-1\right)\times 2^{m-r+2}}\times n.
\end{equation}
The latency of one iteration of the $r$-order IPA decoder is:
\begin{equation} \label{eq:latencyr}
{{t}_{(m,r)}} =  t_{\text{proj}}+t_{(m{-}1,r{-}1)}+t_{\text{PreAgg}}+\left\lceil\frac{d_{r-1}\times \left(2^m-2\right)}{N_{\text{Dec}}}\right\rceil+m,
\end{equation}
where $N_{\text{Dec}}$ is the the number of instantiated decoders for the RM$(m-1,r-1)$ code. In addition, $d_{r-1}$ is the delay that should be considered between inserting the inputs to the $(r-1)$-th level of the decoder, which is dependent on the number of available PUs. The latency of the second-order base IPA decoder, i.e., $t_{(m,2)}$ , is calculated based on \eqref{eq:latency}.  

\section{Simulation and Implementation Results}
\label{sec:result}
\subsection{Simulation results}
\label{sec:sim_res}
In this section, we present simulation results of the decoding performance of the IPA decoder for both hard- and soft-decision decoding including the approximate divider in the aggregation step.
{It is important to note that when we refer to our work as IPA in this section, we specifically mean soft-decision IPA. For hard-decision IPA, we use the term HD-IPA.}
 First, we compare IPA decoding to the baseline RPA algorithm for different codes.
Then, we compare IPA decoding for different numbers of iterations as well as different quantization bit-widths to justify our design choices for the hardware implementation.
Furthermore, we compare IPA decoding of RM codes to SCL decoding of polar codes with the same blocklength and rate. We use the 5G-compatible SCL decoder of \cite{matlab5gnr} with an $11$-bit cyclic redundancy check (CRC).
The target codes for simulations are RM$(6,3)$ and RM$(7,2)$ to cover different orders as well as different rates. For the comparison to polar codes, we consider polar codes with $(n,k)$ pairs of $(64,42)$ and $(128,29)$, as they have the same rate and blocklength to the RM$(6,3)$ and RM$(7,2)$ codes, respectively. All simulations are performed over the AWGN channel.

\begin{figure}[t]
	\centering
\begin{tikzpicture}[spy using outlines={magnification=3,circle,size=1.25cm, black,connect spies}]
		\begin{groupplot}[group style={group name=fer_queries, group size= 2 by 1, horizontal sep=10pt, vertical sep=5pt},
			footnotesize,
			height=.65\columnwidth,  width=.575\columnwidth,
			xlabel=Eb\slash No dB,
			ymode=log,
			tick align=inside,
			grid=both, grid style={gray!30},
			/pgfplots/table/ignore chars={|},
			]

			\nextgroupplot[ylabel= FER, ytick pos=left, y label style={at={(axis description cs:-0.20,.5)},anchor=south},ymin=1e-5, ymax = 1e0,xmin=3.25, xmax=5.25, xtick={3,3.5,4,4.5,5}]

\coordinate (pl1) at (axis cs: 4.25,1e-3);
\coordinate (pl2) at (axis cs: 3.75,1e-4);
\spy on (pl1) in node[fill=white] at (pl2);
		
\addplot[ color=black ,mark=*,mark options={scale=0.7} ] coordinates {
( 3.25, 0.01449822)
( 3.50, 0.00753000)
( 3.75, 0.00417000)
( 4.00, 0.00207000)
( 4.25, 0.00096000)
( 4.50, 0.00045000)
( 4.75, 0.00018700)
( 5.00, 0.00008000)
( 5.25, 0.00002000)
};\label{gp:RPAtanh}

\addplot[ color=red ,mark=o,mark options={scale=0.7} ] coordinates {
( 3.25, 0.01769755)
( 3.50, 0.00926200)
( 3.75, 0.00537499)
( 4.00, 0.00253853)
( 4.25, 0.00114287)
( 4.50, 0.00053000)
( 4.75, 0.00023100)
( 5.00, 0.00010000)
( 5.25, 0.00002400)
};\label{gp:RPAminsum}

\addplot[ color=black ,mark=diamond* ] coordinates {
( 3.25, 0.01614335)
( 3.50, 0.00844000)
( 3.75, 0.00472000)
( 4.00, 0.00226000)
( 4.25, 0.00094000)
( 4.50, 0.00047500)
( 4.75, 0.00019800)
( 5.00, 0.00008500)
( 5.25, 0.00002200)
};

\addplot[ color=red ,mark=diamond ] coordinates {
( 3.25, 0.01849968)
( 3.50, 0.00973672)
( 3.75, 0.00571350)
( 4.00, 0.00264934)
( 4.25, 0.00120013)
( 4.50, 0.00058000)
( 4.75, 0.00025600)
( 5.00, 0.00010400)
( 5.25, 0.00002600)
( 5.50, 0.00000600)
};\label{gp:SIPAminsum}

\addplot[ color=blue ,mark=diamond*,dashed ] coordinates {
( 3.25, 0.34769306)
( 3.50, 0.28319787)
( 3.75, 0.22178358)
( 4.00, 0.16947716)
( 4.25, 0.12678931)
( 4.50, 0.08847287)
( 4.75, 0.06160291)
( 5.00, 0.04074946)
( 5.25, 0.02567816)
( 5.50, 0.01074925)
};

			\coordinate (top) at (rel axis cs:0,1);

\nextgroupplot[yticklabels=\empty,ymin=1e-5, ymax = 1e0,xmin=1.5, xmax=4, xtick={1.5,2,2.5,3,3.5,4}]
\coordinate (bot) at (rel axis cs:1,0);

\coordinate (pr1) at (axis cs: 3,6e-4);
\coordinate (pr2) at (axis cs: 2.25,1e-4);
\spy on (pr1) in node[fill=white] at (pr2);

\addplot[ color=black ,mark=*,mark options={scale=0.7} ] coordinates {
( 1.50, 0.02606984)
( 1.75, 0.01568664)
( 2.00, 0.00888300)
( 2.25, 0.00487500)
( 2.50, 0.00246200)
( 2.75, 0.00120900)
( 3.00, 0.00056900)
( 3.25, 0.00022600)
( 3.50, 0.00009300)
( 3.75, 0.00003700)
( 4.00, 0.00000700)
};

\addplot[ color=red ,mark=o,mark options={scale=0.7} ] coordinates {
( 1.50, 0.02904000)
( 1.75, 0.01763000)
( 2.00, 0.01031000)
( 2.25, 0.00559000)
( 2.50, 0.00298000)
( 2.75, 0.00161000)
( 3.00, 0.00071000)
( 3.25, 0.00027500)
( 3.50, 0.00012000)
( 3.75, 0.00004600)
( 4.00, 0.00001100)
};

\addplot[ color=black ,mark=diamond* ] coordinates {
( 1.50, 0.02606984)
( 1.75, 0.01568664)
( 2.00, 0.00888300)
( 2.25, 0.00487500)
( 2.50, 0.00246200)
( 2.75, 0.00120900)
( 3.00, 0.00056900)
( 3.25, 0.00022600)
( 3.50, 0.00009300)
( 3.75, 0.00003700)
( 4.00, 0.00000700)
};\label{gp:SIPAtanh}

\addplot[ color=red ,mark=diamond ] coordinates {
( 1.50, 0.02904000)
( 1.75, 0.01763000)
( 2.00, 0.01031000)
( 2.25, 0.00559000)
( 2.50, 0.00298000)
( 2.75, 0.00161000)
( 3.00, 0.00071000)
( 3.25, 0.00027500)
( 3.50, 0.00012000)
( 3.75, 0.00004600)
( 4.00, 0.00001100)
};

\addplot[ color=blue ,mark=diamond*,dashed ] coordinates {
( 1.50, 0.30111412)
( 1.75, 0.25779840)
( 2.00, 0.19766752)
( 2.25, 0.15870497)
( 2.50, 0.11010791)
( 2.75, 0.07882084)
( 3.00, 0.05435669)
( 3.25, 0.03265733)
( 3.50, 0.02268191)
( 3.75, 0.01354720)
( 4.00, 0.00797995)
};\label{gp:HIPA}

		\end{groupplot}
		\node[below = 0.8cm of fer_queries c1r1.south] { (a) };
		\node[below = 0.8cm of fer_queries c2r1.south] {(b)};
		\node[below = 1.2cm of fer_queries c1r1.south] {\footnotesize $RM(6,3)$};
		\node[below = 1.2cm of fer_queries c2r1.south] {\footnotesize $RM(7,2)$};		
		\path (top|-current bounding box.north) -- coordinate(legendpos) (bot|-current bounding box.north);
		\matrix[
		matrix of nodes,
		anchor=south,
		draw,
		inner sep=0.1em,
		draw,
		column 1/.style={anchor=base west},
    	column 2/.style={anchor=base west},
    	column 3/.style={anchor=base west},
    	column 4/.style={anchor=base west},
		]at(legendpos)
		{
			\ref{gp:RPAtanh}& \footnotesize RPA (tanh) &[3pt]
			\ref{gp:SIPAtanh}& \footnotesize IPA (tanh)\\
			\ref{gp:RPAminsum}& \footnotesize RPA ($\text{MS}$)&[3pt]
			\ref{gp:SIPAminsum}& \footnotesize IPA ($\text{MS}$)\\
			\ref{gp:HIPA}& \footnotesize HD-IPA \\
			};
	\end{tikzpicture}
	\caption{Comparison of different flavours of RPA and IPA for Reed-Muller codes over AWGN channel.}
	\label{fig:ipa_rpa}
\end{figure}
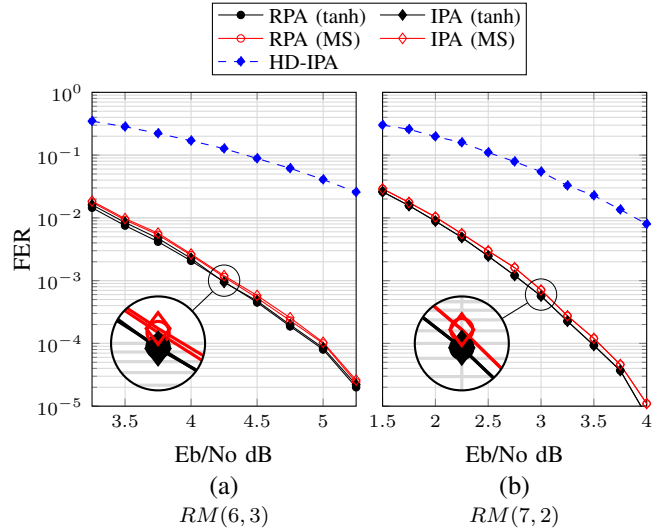
\subsubsection{RPA decoding vs IPA decoding}
Fig.~\ref{fig:ipa_rpa} shows the frame error rate (FER) for RPA and IPA decoding implemented with the exact projection rule \eqref{eq:tanh} and the min-sum approximation \eqref{eq:MinSum}. As in~\cite{Ye2020}, we set the maximum number of iterations $N_{\max} = \lceil {m/2}\rceil$ for the simulations represented in Fig.~\ref{fig:ipa_rpa}.
For second-order RM codes, the only difference between IPA and RPA is the approximate divider during aggregation. The simulation results in Fig.~\ref{fig:ipa_rpa} show that the effect of this approximate divider is negligible as there is effectively no performance difference between IPA and RPA  for the RM$(7,2)$ code.
For the RM$(6,3)$ code, the performance difference between IPA and RPA is also negligible, even though $\sim 567$ internal iterations are removed by the IPA algorithm. 
In addition, Fig.~\ref{fig:ipa_rpa} shows that the degradation caused by the min-sum update rule \eqref{eq:MinSum} compared to the exact update rule \eqref{eq:tanh} is small for both codes.
{In addition, we aim to compare the hardware implementation of our proposed IPA to the fully-parallel  HD-IPA architecture proposed in \cite{Hashemipour-Nazari2021}. Therefore, a comparison between the HD-IPA \cite{Hashemipour-Nazari2021} and IPA is made in 
Fig.~\ref{fig:ipa_rpa}.
As expected, the error-correcting performance of the HD-IPA shows a noticeable degradation of more than 2 dB and 1.75 dB for RM(6,3) and RM(7,3) codes, respectively, in comparison to the soft-input IPA.}

\begin{figure}[t]
	\centering
\begin{tikzpicture}[spy using outlines={magnification=3,circle,size=1.25cm, black,connect spies}]
		\begin{groupplot}[group style={group name=fer_queries, group size= 2 by 1, horizontal sep=10pt, vertical sep=5pt},
			footnotesize,
			height=.65\columnwidth,  width=.575\columnwidth,
			xlabel=Eb\slash No dB,
			ymode=log,
			tick align=inside,
			grid=both, grid style={gray!30},
			/pgfplots/table/ignore chars={|},
			]

\nextgroupplot[ylabel= FER, ytick pos=left, y label style={at={(axis description cs:-0.20,.5)},anchor=south},ymin=1e-5, ymax = 1e-1,xmin=3.25, xmax=5.25, xtick={3,3.5,4,4.5,5,5.5}]
			
\coordinate (pl1) at (axis cs: 4.25,1.2e-3);
\coordinate (pl2) at (axis cs: 3.75,1e-4);
\spy on (pl1) in node[fill=white] at (pl2);

\addplot[ color=blue ,mark=o,mark options={scale=0.7} ] coordinates {
( 3.25, 0.01849968)
( 3.50, 0.00973672)
( 3.75, 0.00571350)
( 4.00, 0.00264934)
( 4.25, 0.00120013)
( 4.50, 0.00058000)
( 4.75, 0.00025600)
( 5.00, 0.00010400)
( 5.25, 0.00002500)
( 5.50, 0.00000600)
};\label{gp:SIPAminsum_nmax3}

\addplot[ color=dgreen ,mark=triangle ] coordinates {
( 3.25, 0.06128202)
( 3.50, 0.03512593)
( 3.75, 0.02074517)
( 4.00, 0.01084622)
( 4.25, 0.00521121)
( 4.50, 0.00233471)
( 4.75, 0.00106301)
( 5.00, 0.00040700)
( 5.25, 0.00013800)
( 5.50, 0.00003600)
};\label{gp:SIPAminsum_nmax1}

\addplot[ color=red ,mark=square,mark options={scale=0.7} ] coordinates {
( 3.25, 0.01905451)
( 3.50, 0.00996780)
( 3.75, 0.00584819)
( 4.00, 0.00274021)
( 4.25, 0.00122441)
( 4.50, 0.00058700)
( 4.75, 0.00025600)
( 5.00, 0.00010400)
( 5.25, 0.00002600)
( 5.50, 0.00000600)
};\label{gp:SIPAminsum_nmax2}

			\coordinate (top) at (rel axis cs:0,1);

\nextgroupplot[yticklabels=\empty,ymin=1e-5, ymax = 1e-1,xmin=1.5, xmax=4, xtick={1.5,2,2.5,3,3.5,4}]

\coordinate (bot) at (rel axis cs:1,0);

\coordinate (pr1) at (axis cs: 3,7e-4);
\coordinate (pr2) at (axis cs: 2.25,1e-4);
\spy on (pr1) in node[fill=white] at (pr2);

\addplot[ color=blue ,mark=o,mark options={scale=0.7}] coordinates {
( 1.50, 0.02988286)
( 1.75, 0.01813697)
( 2.00, 0.01045577)
( 2.25, 0.00561864)
( 2.50, 0.00299659)
( 2.75, 0.00161975)
( 3.00, 0.00072500)
( 3.25, 0.00028500)
( 3.50, 0.0001200)
( 3.75, 0.00004200)
( 4.00, 0.00001100)
};

\addplot[ color=red ,mark=square,mark options={scale=0.7} ] coordinates {
( 1.50, 0.03383000)
( 1.75, 0.02045000)
( 2.00, 0.01190000)
( 2.25, 0.00634000)
( 2.50, 0.00330000)
( 2.75, 0.00177000)
( 3.00, 0.00082000)
( 3.25, 0.00031100)
( 3.50, 0.00012600)
( 3.75, 0.00004900)
( 4.00, 0.00001200)
};

\addplot[ color=dgreen ,mark=triangle ] coordinates {
( 1.50, 0.21097937)
( 1.75, 0.15229509)
( 2.00, 0.10320556)
( 2.25, 0.06636000)
( 2.50, 0.04124000)
( 2.75, 0.02310000)
( 3.00, 0.01257000)
( 3.25, 0.00570434)
( 3.50, 0.00253993)
( 3.75, 0.00108949)
( 4.00, 0.00043100)
};

\addplot[ color=black ,mark=pentagon ] coordinates {
( 1.50, 0.02904000)
( 1.75, 0.01763000)
( 2.00, 0.01031000)
( 2.25, 0.00559000)
( 2.50, 0.00298000)
( 2.75, 0.00161000)
( 3.00, 0.00071000)
( 3.25, 0.00028500)
( 3.50, 0.00012000)
( 3.75, 0.00004100)
( 4.00, 0.00001100)
};\label{gp:SIPAminsum_nmax4}

		\end{groupplot}
		\node[below = 0.8cm of fer_queries c1r1.south] { (a) };
		\node[below = 0.8cm of fer_queries c2r1.south] {(b)};
		\node[below = 1.2cm of fer_queries c1r1.south] {\footnotesize $RM(6,3)$ };
		\node[below = 1.2cm of fer_queries c2r1.south] {\footnotesize $RM(7,2)$ };		
		\path (top|-current bounding box.north) -- coordinate(legendpos) (bot|-current bounding box.north);
		\matrix[
		matrix of nodes,
		anchor=south,
		draw,
		inner sep=0.1em,
		draw,
		column 1/.style={anchor=base west},
    	column 2/.style={anchor=base west},
    	column 3/.style={anchor=base west},
    	column 4/.style={anchor=base west},
		]at(legendpos)
		{
			\ref{gp:SIPAminsum_nmax1}& \footnotesize 1 Iterations &[3pt]
			\ref{gp:SIPAminsum_nmax3}& \footnotesize 3 Iterations \\
			\ref{gp:SIPAminsum_nmax2}& \footnotesize 2 Iterations&[3pt]
			\ref{gp:SIPAminsum_nmax4}& \footnotesize 4 Iteration \\			
			};
	\end{tikzpicture}
	\caption{Comparison of different numbers of iterations for the IPA algorithm.}
	\label{fig:ipa_itr}
\end{figure}
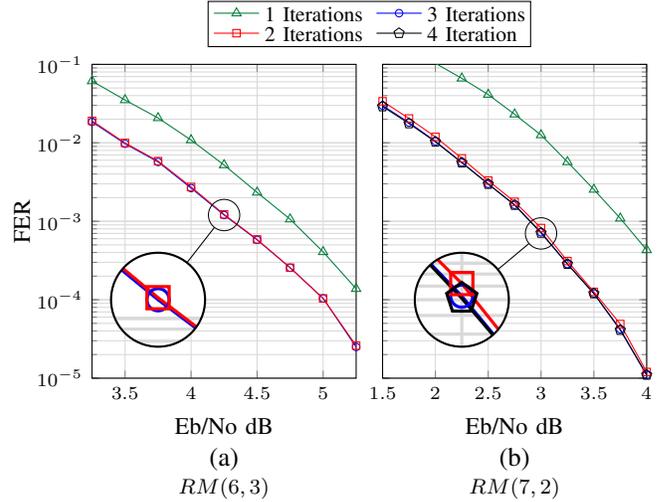
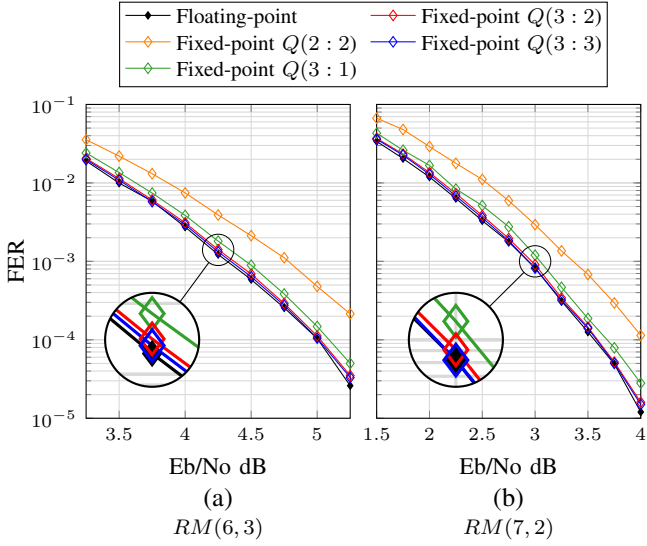
\begin{figure}[t]
	\centering
\begin{tikzpicture}[spy using outlines={magnification=3,circle,size=1.25cm, black,connect spies}]
		\begin{groupplot}[group style={group name=fer_queries, group size= 2 by 1, horizontal sep=10pt, vertical sep=5pt},
			footnotesize,
			height=.65\columnwidth,  width=.575\columnwidth,
			xlabel=Eb\slash No dB,
			ymode=log,
			tick align=inside,
			grid=both, grid style={gray!30},
			/pgfplots/table/ignore chars={|},
			]

			\nextgroupplot[ylabel= FER, ytick pos=left, y label style={at={(axis description cs:-0.20,.5)},anchor=south},ymin=1e-5, ymax = 1e-1,xmin=3.25, xmax=5.25, xtick={2.5,3,3.5,4,4.5,5}]

\coordinate (pl1) at (axis cs: 4.25,1.4e-3);
\coordinate (pl2) at (axis cs: 3.75,1e-4);
\spy on (pl1) in node[fill=white] at (pl2);

\addplot[ color=black ,mark=diamond*,mark options={scale=0.7} ] coordinates {
( 3.25, 0.01905451)
( 3.50, 0.00996780)
( 3.75, 0.00584819)
( 4.00, 0.00274021)
( 4.25, 0.00122441)
( 4.50, 0.00058700)
( 4.75, 0.00025600)
( 5.00, 0.00010400)
( 5.25, 0.00002600)
( 5.50, 0.00000600)
};\label{gp:float}

\addplot[ color=Paired-7 ,mark=diamond ] coordinates {
( 3.25, 0.03554570)
( 3.50, 0.02194445)
( 3.75, 0.01310819)
( 4.00, 0.00746000)
( 4.25, 0.00389800)
( 4.50, 0.00212500)
( 4.75, 0.00111100)
( 5.00, 0.00047900)
( 5.25, 0.00021400)
( 5.50, 0.00008700)
};

\addplot[ color=Paired-3 ,mark=diamond ] coordinates {
( 3.25, 0.02408060)
( 3.50, 0.01354166)
( 3.75, 0.00742500)
( 4.00, 0.00385700)
( 4.25, 0.00181000)
( 4.50, 0.00089400)
( 4.75, 0.00038400)
( 5.00, 0.00014800)
( 5.25, 0.00005000)
( 5.50, 0.00001500)
};

\addplot[ color=red ,mark=diamond ] coordinates {
( 3.25, 0.02052507)
( 3.50, 0.01136313)
( 3.75, 0.00606600)
( 4.00, 0.00310400)
( 4.25, 0.00141200)
( 4.50, 0.00068900)
( 4.75, 0.00029100)
( 5.00, 0.00011200)
( 5.25, 0.00003500)
( 5.50, 0.00000800)
};\label{gp:fixed32}

\addplot[ color=blue ,mark=diamond ] coordinates {
( 3.25, 0.01983714)
( 3.50, 0.01083641)
( 3.75, 0.00576200)
( 4.00, 0.00293600)
( 4.25, 0.00132300)
( 4.50, 0.00063700)
( 4.75, 0.00027500)
( 5.00, 0.00010500)
( 5.25, 0.00003300)
( 5.50, 0.00000600)
};

			\coordinate (top) at (rel axis cs:0,1);

\nextgroupplot[yticklabels=\empty,ymin=1e-5, ymax = 1e-1,xmin=1.5, xmax=4, xtick={1.5,2,2.5,3,3.5,4}]

\coordinate (bot) at (rel axis cs:1,0);

\coordinate (pr1) at (axis cs: 3,1e-3);
\coordinate (pr2) at (axis cs: 2.25,1e-4);
\spy on (pr1) in node[fill=white] at (pr2);

\addplot[ color=black ,mark=diamond*,mark options={scale=0.7} ] coordinates {
( 1.50, 0.03383000)
( 1.75, 0.02045000)
( 2.00, 0.01190000)
( 2.25, 0.00634000)
( 2.50, 0.00330000)
( 2.75, 0.00177000)
( 3.00, 0.00082000)
( 3.25, 0.00031100)
( 3.50, 0.00012600)
( 3.75, 0.00004900)
( 4.00, 0.00001200)
};

\addplot[ color=Paired-7 ,mark=diamond ] coordinates {
( 1.50, 0.06685833)
( 1.75, 0.04787438)
( 2.00, 0.02907653)
( 2.25, 0.01776672)
( 2.50, 0.01107493)
( 2.75, 0.00593014)
( 3.00, 0.00293778)
( 3.25, 0.00135781)
( 3.50, 0.00068000)
( 3.75, 0.00029500)
( 4.00, 0.00011400)
};\label{gp:fixed22}

\addplot[ color=Paired-3 ,mark=diamond ] coordinates {
( 1.50, 0.04301075)
( 1.75, 0.02610285)
( 2.00, 0.01672520)
( 2.25, 0.00834585)
( 2.50, 0.00514271)
( 2.75, 0.00276228)
( 3.00, 0.00119837)
( 3.25, 0.00046700)
( 3.50, 0.00018700)
( 3.75, 0.00007900)
( 4.00, 0.00002800)
};\label{gp:fixed31}

\addplot[ color=red ,mark=diamond ] coordinates {
( 1.50, 0.03700962)
( 1.75, 0.02334104)
( 2.00, 0.01354646)
( 2.25, 0.00731149)
( 2.50, 0.00383864)
( 2.75, 0.00195996)
( 3.00, 0.00090700)
( 3.25, 0.00034200)
( 3.50, 0.00014500)
( 3.75, 0.00005200)
( 4.00, 0.00001600)
};

\addplot[ color=blue ,mark=diamond ] coordinates {
( 1.50, 0.03599712)
( 1.75, 0.02255402)
( 2.00, 0.01288876)
( 2.25, 0.00680342)
( 2.50, 0.00359614)
( 2.75, 0.00182827)
( 3.00, 0.00082000)
( 3.25, 0.00032100)
( 3.50, 0.00014000)
( 3.75, 0.00005000)
( 4.00, 0.00001500)
};\label{gp:fixed33}

		\end{groupplot}
		\node[below = 0.8cm of fer_queries c1r1.south] { (a) };
		\node[below = 0.8cm of fer_queries c2r1.south] {(b)};
		\node[below = 1.2cm of fer_queries c1r1.south] {\footnotesize $RM(6,3)$};
		\node[below = 1.2cm of fer_queries c2r1.south] {\footnotesize $RM(7,2)$};		
		\path (top|-current bounding box.north) -- coordinate(legendpos) (bot|-current bounding box.north);
		\matrix[
		matrix of nodes,
		anchor=south,
		draw,
		inner sep=0.1em,
		draw,
		column 1/.style={anchor=base west},
    	column 2/.style={anchor=base west},
		]at(legendpos)
		{
			\ref{gp:float}  & \footnotesize Floating-point &[3pt]
			\ref{gp:fixed32}& \footnotesize Fixed-point $Q(3:2$) \\
			\ref{gp:fixed22}& \footnotesize Fixed-point $Q(2:2$) &[3pt]			
			\ref{gp:fixed33}& \footnotesize Fixed-point $Q(3:3$)  \\
			\ref{gp:fixed31}& \footnotesize Fixed-point $Q(3:1$) \\
			};
	\end{tikzpicture}
	\caption{FER comparison between floating-point and different fixed-point implementation of IPA decoding.}
	\label{fig:fixed_float}
\end{figure}

\subsubsection{Number of iterations}
The hardware cost of our architecture represented in Fig.~\ref{fig:SIPA_3rd} scales linearly with the number of iterations. Therefore, we also explored whether the number of iterations can be reduced without degrading the error-correcting performance significantly. Fig.~\ref{fig:ipa_itr} shows the performance of IPA decoding with $1$ to $ \lceil {m/2}\rceil$ iterations. We observe that the performance loss with one iteration is significant compared to $\lceil {m/2}\rceil$ iterations for both examined RM codes. However, there is no performance degradation for the RM$(6,3)$ code and only a very small degradation for RM$(7,3)$ with $2$ iterations, instead of $\lceil {m/2}\rceil = 3$ and $\lceil {m/2}\rceil = 4$ iterations, respectively. Therefore, we set $N_{\max} = 2$ for the hardware implementation of the IPA decoder for both of these codes, reducing the required hardware and the latency by $33\%$ and $50\%$ for each code, respectively.

\subsubsection{Quantization bit-width}
For the fixed-point implementation, we quantized all LLRs using a $Q(q_i:q_f)$ quantization scheme, where $q_i$ and $q_f$ is the number of integer and fractional bits, respectively. We increase the bit-width by one at each stage of the FHT inside the FOD because it is very sensitive to saturation. The results of all other operations are always clipped to remain within the representable range. Fig.~\ref{fig:fixed_float} shows the results for the different values of $q_i$ and $q_f$. We observe that the $5$-bit quantized IPA decoder with $q_i=3$ and $q_f=2$ is almost as accurate as the floating-point IPA decoder for both considered codes. Therefore, we present hardware implementation results using $5$-bit LLRs.
 
\begin{figure}[t]
	\centering
\begin{tikzpicture}[spy using outlines={magnification=3,circle,size=1.25cm, black,connect spies}]
		\begin{groupplot}[group style={group name=fer_queries, group size= 2 by 1, horizontal sep=10pt, vertical sep=5pt},
			footnotesize,
			height=.65\columnwidth,  width=.575\columnwidth,
			xlabel=Eb\slash No dB,
			ymode=log,
			tick align=inside,
			grid=both, grid style={gray!30},
			/pgfplots/table/ignore chars={|},
			]

			\nextgroupplot[ylabel= FER, ytick pos=left, y label style={at={(axis description cs:-0.20,.5)},anchor=south},ymin=1e-5, ymax = 1e-1,xmin=3.25, xmax=5.25, xtick={2.5,3,3.5,4,4.5,5}]

\addplot[ color=black ,mark=diamond* ] coordinates {
( 3.25, 0.02052507)
( 3.50, 0.01136313)
( 3.75, 0.00606600)
( 4.00, 0.00310400)
( 4.25, 0.00141200)
( 4.50, 0.00068900)
( 4.75, 0.00029100)
( 5.00, 0.00011200)
( 5.25, 0.00003500)
( 5.50, 0.00000800)
};\label{gp:quant32}

\addplot[ color=dgreen ,mark=pentagon ] coordinates {
( 3.25, 0.05451374)
( 3.50, 0.03326680)
( 3.75, 0.01769191)
( 4.00, 0.00885183)
( 4.25, 0.00434582)
( 4.50, 0.00189399)
( 4.75, 0.00079800)
( 5.00, 0.00029000)
( 5.25, 0.00010500)
};\label{gp:scl32}

\addplot[ color=dgreen ,mark=triangle ] coordinates {
( 3.25, 0.08125457)
( 3.50, 0.05296330)
( 3.75, 0.02964544)
( 4.00, 0.01706659)
( 4.25, 0.00845173)
( 4.50, 0.00412150)
( 4.75, 0.00195578)
( 5.00, 0.00076100)
( 5.25, 0.00029500)
};\label{gp:scl16}

\addplot[ color=dgreen ,mark=star ] coordinates {
( 3.25, 0.03554671)
( 3.50, 0.01994575)
( 3.75, 0.01046704)
( 4.00, 0.00537355)
( 4.25, 0.00251393)
( 4.50, 0.00108300)
( 4.75, 0.00047000)
( 5.00, 0.00017600)
( 5.25, 0.00005500)
};\label{gp:scl64}

			\coordinate (top) at (rel axis cs:0,1);

\nextgroupplot[yticklabels=\empty,ymin=1e-5, ymax = 1e-1,xmin=1.5, xmax=4, xtick={1.5,2,2.5,3,3.5,4}]
\coordinate (bot) at (rel axis cs:1,0);

\coordinate (pr1) at (axis cs: 3,8e-4);
\coordinate (pr2) at (axis cs: 2.25,1e-4);
\spy on (pr1) in node[fill=white] at (pr2);

\addplot[ color=black ,mark=diamond* ] coordinates {
( 1.50, 0.03700962)
( 1.75, 0.02334104)
( 2.00, 0.01354646)
( 2.25, 0.00731149)
( 2.50, 0.00383864)
( 2.75, 0.00195996)
( 3.00, 0.00090700)
( 3.25, 0.00034200)
( 3.50, 0.00014500)
( 3.75, 0.00005200)
( 4.00, 0.00001600)
};

\addplot[ color=dgreen ,mark=pentagon ] coordinates {
( 1.50, 0.05754000)
( 1.75, 0.03632000)
( 2.00, 0.02022000)
( 2.25, 0.01170000)
( 2.50, 0.00575000)
( 2.75, 0.00258000)
( 3.00, 0.00113000)
( 3.25, 0.00042700)
( 3.50, 0.00017200)
( 3.75, 0.00005700)
( 4.00, 0.00001900)
};

\addplot[ color=dgreen ,mark=triangle ] coordinates {
( 1.50, 0.08235000)
( 1.75, 0.05427000)
( 2.00, 0.03249000)
( 2.25, 0.01936000)
( 2.50, 0.01027000)
( 2.75, 0.00504000)
( 3.00, 0.00237000)
( 3.25, 0.00100600)
( 3.50, 0.00042000)
( 3.75, 0.00015800)
( 4.00, 0.00005900)
};

\addplot[ color=dgreen ,mark=star ] coordinates {
( 1.50, 0.04103355)
( 1.75, 0.02472231)
( 2.00, 0.01340580)
( 2.25, 0.00683000)
( 2.50, 0.00331600)
( 2.75, 0.00144400)
( 3.00, 0.00059100)
( 3.25, 0.00020500)
( 3.50, 0.00008100)
( 3.75, 0.00003000)
( 4.00, 0.00000700)
};

		\end{groupplot}
		\node[below = 0.8cm of fer_queries c1r1.south] { (a) };
		\node[below = 0.8cm of fer_queries c2r1.south] {(b)};
		\node[below = 1.2cm of fer_queries c1r1.south] {\footnotesize $RM(6,3)$ and $Polar(64,42)$ };
		\node[below = 1.2cm of fer_queries c2r1.south] {\footnotesize $RM(7,2)$ and $Polar(128,29)$};		
		\path (top|-current bounding box.north) -- coordinate(legendpos) (bot|-current bounding box.north);
		\matrix[
		matrix of nodes,
		anchor=south,
		draw,
		inner sep=0.1em,
		draw,
		column 1/.style={anchor=base west},
    	column 2/.style={anchor=base west},
		]at(legendpos)
		{
			\ref{gp:quant32}& \footnotesize IPA (Q($3:2$)) &[3pt]
			\ref{gp:scl32}& \footnotesize SCL (L = $32$) \\			
			\ref{gp:scl16}& \footnotesize SCL (L = $16$)  &[3pt]
			\ref{gp:scl64}& \footnotesize SCL (L = $64$) \\
			};
	\end{tikzpicture}
	\caption{FER comparison between 5-bit fixed-point IPA decoding and floating-point SCL decoding.}
	\label{fig:ipa_polar}
\end{figure}
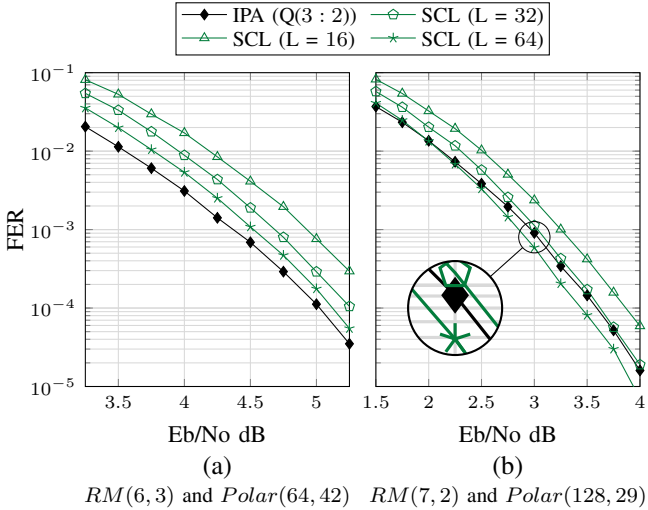
\subsubsection{IPA decoding of RM codes vs SCL decoding of polar codes}
For the final comparison, we compare the performance of our selected $5$-bit quantized IPA with $2$ iterations to the {floating-point} SCL decoding of 5G polar codes with $11$-bit CRC. 
As Fig.~\ref{fig:ipa_polar} demonstrates, the quantized IPA decoder for RM$(6,3)$ outperforms the floating-point SCL decoder for a polar code of the same blocklength and rate with list sizes of $L=16$, $L=32$, and $L=64$. Moreover, the quantized IPA decoder for RM$(7,2)$ outperforms the floating-point SCL decoder for a polar code of the same blocklength and rate with list size of $L=16$ and $L=32$, but has worse performance for $L=64$. 


\subsection{ASIC synthesis results}
\label{sec:impl_res}

In this section, we present synthesis results for our proposed IPA decoder architecture. The IPA decoder has been implemented in VHDL and synthesized using the Cadence Genus RTL compiler with the STM $28\text{nm}$ FD-SOI technology in the slow-slow corner and at $25^\text{o}$~C.
{ To obtain accurate power measurements, we performed gate-level (GL) simulations by generating a standard delay file (SDF) through synthesis.
Subsequently, we utilized the SDF to perform GL simulations in the Cadence Xcelium simulator. 
To ensure an accurate power estimate, we incorporated the switching activity obtained from the GL simulation for $10^3$ frames into our analysis.}
We show the synthesis results for the IPA decoder for RM$(7,2)$ and RM$(6,3)$ codes. 
As concluded from Fig.~\ref{fig:fixed_float}, the IPA decoder has been implemented for $5$-bit input LLRs, i.e., $Q(3:2)$ and two decoding iterations. 
Furthermore, we provide the synthesis results for different numbers of PUs employed in the IPA decoder to show the trade-off between area consumption and latency as well as throughput.  

Since, to the best of our knowledge, there are no other hardware implementations of {soft-decision} projection-aggregation based RM decoders in the literature, we compare the area consumption, latency, and throughput of the IPA decoder against the state-of-the-art SCL decoder of~\cite{Ren2022} synthesized for polar codes with the same blocklength and information rate and with the same technology and settings.
{The SCL decoder has been synthesized for $6$-bit quantization to ensure that there is minimal performance loss between the SCL decoder with quantized LLRs and the floating-point LLRs\cite{Ren2022}.}
Additionally, we select the list sizes based on Fig.~\ref{fig:ipa_polar} and we allow the SCL decoder to have slightly worse error-correcting performance if necessary to be as conservative in our comparison as possible.
{ We also synthesized the fully-parallel HD-IPA decoder\cite{Hashemipour-Nazari2021} for comparison.
To ensure a fair comparison with the soft-decision IPA, we also synthesized the HD-IPA decoder for 2 iterations.}
\begin{table}[t]
\captionsetup{textfont={sc,footnotesize},justification=centerlast, labelsep=newline}
\centering
\caption{\label{tab:rm_6_3} Synthesis results for our proposed IPA decoder for RM$(6,3)$ code and for the state-of-the-art SCL decoder of~\cite{Ren2022} for a 5G polar code with $n=64$ and $k=42$.}

\begin{adjustbox}{width=\columnwidth,center}

\begin{threeparttable}
\begin{tabular}{lccccc}
\toprule
\textbf{Code}                 & \multicolumn{4}{c}{RM$(6,3)$}  & \multicolumn{1}{c}{Polar $(64,42)$} \\ 
\textbf{Decoder}              & \multicolumn{3}{c}{IPA}  & \multicolumn{1}{c}{HD-IPA\cite{Hashemipour-Nazari2021}} & \multicolumn{1}{c}{SCL\cite{Ren2022}}  \\ 
\midrule
\textbf{List size $(\mathrm{L})$}            & - & - & - & {-} &  $64$ \\ 
\textbf{Number of PUs $(\mathrm{P})$}        & $16$  & $32$ & $64$ & {$8001$\tnote{a}}  & - \\ 
\midrule
\textbf{Clock Rate $(\mathrm{MHz})$}    & $714$ & $714$ & $714$ & {$500$} &$350$ \\ 
\textbf{Latency  $(\mathrm{cc})$ } & $294$ & $168$ & $106$ & {$26$} &$95$ \\ 
\textbf{Latency $(\mathrm{\mu s})$ } & $0.411$ & $0.235$ & $0.148$ & {$0.052$} &$0.270$ \\
\textbf{Throughput $(\mathrm{Mbps})$}    & $357$ & $714$ & $1428$ & {$1231$} &$235$  \\ 
\midrule
\textbf{Area ($\left(\mathrm{mm}^2\right)$}        & $0.38$ & $0.61$ & $1.21$ & {{${4.08}$}} &$2.04$ \\
\textbf{Area Eff. $\left(\mathrm{Gbps} / \mathrm{mm}^2\right)$} & $0.94$ & $1.17$ & $1.60 $ & {$0.30$} &$0.12$ \\
\midrule
{\textbf{Power $(\mathrm{mW})$}} & {$317$} & {$505$} & {$801$} & {NA\tnote{b}} & {NA\tnote{b}} \\
{\textbf{Energy $(\mathrm{pJ} / \mathrm{b})$}} & {$888$} & {$707$} &{$561$} & {NA}& {NA} \\
\bottomrule
\end{tabular}

{
\begin{tablenotes}[flushleft]
\item[a]There is no PU defined in \cite{Hashemipour-Nazari2021}, but there are $127\times 63$ parallel units including two levels of projection, FOD, and two levels of aggregation.
\item[b] Not available: Due to the high resource utilization, the post-synthesis power analysis failed to run.
\end{tablenotes}
}

\end{threeparttable}

\end{adjustbox}

\end{table}
\begin{table}[t]
\captionsetup{textfont={sc,footnotesize},justification=centerlast, labelsep=newline}
\centering
\caption{Synthesis results for our proposed IPA decoder for RM$(7,2)$ code and for the state-of-the-art SCL decoder of~\cite{Ren2022} for a 5G polar code with $n=128$ and $k=29$.}
\label{tab:rm7_2}
\begin{adjustbox}{width=\columnwidth,center}
\begin{tabular}{lccccc}
\toprule
\textbf{Code}                 & \multicolumn{4}{c}{RM$(7,2)$}  & \multicolumn{1}{c}{Polar $(128,29)$} \\ 
\textbf{Decoder}              & \multicolumn{3}{c}{IPA}  & \multicolumn{1}{c}{HD-IPA\cite{Hashemipour-Nazari2021}} &  \multicolumn{1}{c}{SCL\cite{Ren2022} }  \\ 
\midrule
\textbf{List size $(\mathrm{L})$}            & - & - & - & {-} & $32$ \\ 
\textbf{Number of PUs $(\mathrm{P})$}        & $2$  & $4$ & $8$  & {$127$} & - \\ 
\midrule
\textbf{Clock Rate $(\mathrm{MHz})$}    & $555$ & $555$ & $555$ & {$384$} & $403$ \\ 
\textbf{Latency $(\mathrm{cc})$ } & $156$ & $92$ & $60$ & {$15$} &$94$ \\ 
\textbf{Latency $(\mathrm{\mu s})$ } & $0.281$ & $0.165$ & $0.108$ & {$0.039$} &$0.233$ \\
\textbf{Throughput $(\mathrm{Mbps})$}    & $1110$ & $2220$ & $4440$ & {$3282$} &$548$  \\ 
\midrule
\textbf{Area $\left(\mathrm{mm}^2\right)$}        & $0.33$ & $0.50$ & $0.88$ & {$1.90$} &$0.96$ \\
\textbf{Area Eff. $\left(\mathrm{Gbps} / \mathrm{mm}^2\right)$} & $3.30$ & $4.35$ & $5.04$ & {$1.72$} &$0.57$ \\
\midrule
{\textbf{Power $(\mathrm{mW})$}} & {$214$} & {$364$} & {$662$} &{$237$} &{$52$} \\
{\textbf{Energy $(\mathrm{pJ} / \mathrm{b})$}} & {$192$} & {$163$} & {$149$} & {$72$} &{$95$} \\
\bottomrule
\end{tabular}
\end{adjustbox}

\end{table}

Table~\ref{tab:rm_6_3} presents the synthesis results for soft-decision IPA decoders { with various numbers of PUs and fully-parallel HD-IPA for RM$(6,3)$ code. It also compares the IPA decoders with the SCL decoder of \cite{Ren2022} for the 5G polar code with $n=64$ and $k=42$, which has the same rate and blocklenth as the RM code. 
The second-order decoder instantiated in the IPA decoder for the RM$(6,3)$ code performs $32$ projections, resulting in $32$ PUs in case of a fully-parallel implementation. Therefore, we synthesized the IPA decoder for $16$ and $32$ PUs to show the effects of the partially-parallel and fully-parallel second-order decoder used in the IPA decoder for decoding a third-order code. In addition, Table~\ref{tab:rm_6_3} also includes synthesis results for $64$ PUs, meaning that two fully-parallel second-order decoders are employed in the third-order IPA decoder.}
The results show that increasing the number of PUs from $16$ to $32$ results in a $58\%$ increase in the area while the latency decreases by $42\%$. However, implementing the IPA decoder with $64$ PUs results in slightly higher than two times area consumption compared to $32$ PUs since there are two decoders with separate control units and memories instantiated in the third-order IPA decoder. 
{ Furthermore, we observed that HD-IPA shows significantly lower latency compared to IPA due to its fully-parallel architecture.
However, it has a significantly worse error-correcting performance, as shown in Fig.~\ref{fig:ipa_rpa}. Moreover, this remarkably low latency is achieved at the cost of using significantly more resources, making it impractical for many applications.
This further emphasizes the necessity for a partial-parallel architecture, such as our proposed architecture, which offers a more viable alternative.
Additionally, despite its lower latency, HD-IPA does not achieve high throughput due to its non-pipelined structure, resulting in very low area efficiency.}
We also see that {all different configurations for the proposed IPA decoder} have a significantly smaller area and a much higher throughput than the SCL decoder, resulting in an improvement in the area efficiency of one order of magnitude. At the same time, the IPA decoders have $13$\% and $45$\% lower absolute latency than the SCL decoder for $P=32$ and $P=64$, respectively.


Table~\ref{tab:rm7_2} presents the synthesis results for various soft-decision IPA decoders and  {HD-IPA} for the RM$(7,2)$ code as well as an SCL decoder  for the 5G polar code with $n=128$ and $k=29$. 
{We see that HD-IPA shows up to 7 times lower latency but requires up to 6 times more resources.
Similarly to what was observed in Table~\ref{tab:rm_6_3}, HD-IPA is less efficient in terms of resource usage compared to IPA because it does not use a pipelined architecture.
However, the resource efficiency for RM$(7,2)$ is better than RM$(6,3)$ due to the lower number of clock cycles required for decoding each codeword.
It is again important to highlight that the error-correcting performance of HD-IPA is significantly worse than that of the soft-decision IPA as it is shown in Fig. \ref{fig:ipa_rpa}.
}
Similarly to the previous results, all IPA decoders have a significantly smaller area and a much higher throughput than the SCL decoder, resulting in an improvement in the area efficiency by a factor between $6$ and $9$, depending on the number of PUs. At the same time, the IPA decoders have $29$\% and $54$\% lower absolute latency than the SCL decoder for $P=4$ and $P=8$, respectively.

{Table~\ref{tab:rm_6_3} and Table~\ref{tab:rm7_2} also provide a comparison of the power consumption among the soft-decision IPA, HD-IPA, and SCL decoder. The proposed IPA architecture shows higher power consumption compared to the other two decoders. This can be attributed to its pipelined architecture, where all components remain active during each clock cycle, contrary to the SCL decoder that mostly consists of memory.
Furthermore, it is evident that as the level of parallelism, indicated by the number of PUs, increases the energy consumption per bit decreases. This implies that the IPA architecture can be configured to deliver a high-throughput decoder with reasonable energy consumption per bit if such performance is required.

We have included detailed information regarding the average area per iteration and the power consumption for each iteration of any block in the IPA decoder in Table~\ref{tab:ap_break}.
As expected, the PU utilizes the largest area among all blocks. Additionally, within the PU, the \textit{FOD} block is the most area-intensive component. The area consumption of the register array, designed to fulfill pipeline requirements, is negligible compared to other blocks, as it only needs to store a small number of codewords.
Furthermore, it is worth noting that the area utilization for one PU in the second-order decoder embedded within the decoder for RM$(6,3)$ is relatively low. This is because we employed a fully-parallel second-order decoder with $32$ PUs at that level, resulting in smaller \textit{Projection} and \textit{PreAggregation} components.
Similarly, the divider at that level appears relatively large compared to other parts, as it has $32$ vectors ready at each clock cycle, requiring five levels of parallel adders and shift registers, as depicted in Fig.~\ref{fig:divider}.

In terms of power consumption, we reported the power values for both iterations. The second iteration is less computationally intensive, as the majority of error correction occurs in the first iteration. Consequently, the \textit{FOD} units exhibit lower activity levels during the second iteration. Therefore, we observe approximately $18\%$ and $14\%$ less power consumption for the second iteration of the IPA decoder for RM$(7,2)$ and RM$(6,3)$ codes, respectively.
Additionally, we noticed that the power consumption of the control units, register array, and divider for the second iteration remains almost the same as the first iteration, as their activities are not dependent on input values.}
\begin{table}[t]
\captionsetup{textfont={sc,footnotesize},justification=centerlast, labelsep=newline}
\centering
\caption{Area utilization and power consumption of each component of the IPA decoder}
\label{tab:ap_break}

\begin{adjustbox}{width=\columnwidth,center}

\begin{threeparttable}
{\begin{tabular}{lcccccc}
\toprule
 \multicolumn{1}{c}{} & \multicolumn{3}{c}{\textbf{RM$(7,2)$, $\mathrm{P}=4$}} & \multicolumn{3}{c}{\textbf{RM$(6,3)$, $\mathrm{P}=32$}} \\
 

 \multicolumn{1}{c}{} & \multicolumn{1}{c}{\textbf{Area}\tnote{a}} & \multicolumn{2}{c}{\textbf{Power{($m$W)}}} & \multicolumn{1}{c}{\textbf{Area}\tnote{a}} &\multicolumn{2}{c}{\textbf{Power{($m$W)}}}  \\

\multicolumn{1}{c}{} & \multicolumn{1}{c}{ $\mathrm{mm}^2$} & \multicolumn{1}{c}{\textbf{itr. 1}} &\multicolumn{1}{c}{\textbf{itr. 2}}  &\multicolumn{1}{c}{$\mathrm{mm}^2$}&\multicolumn{1}{c}{\textbf{itr. 1}} &\multicolumn{1}{c}{\textbf{itr. 2}} \\

\midrule
  \textbf{IPA($r=3$)} & - & - &-&$0.306$ & $271.76$ & $232.84$ \\
 \hspace{2mm} \textbf{\textit{Projection}}($r=3$)         &  -                & -         & -         &  $0.007$        & $ \phantom{00}5.40$   & $ \phantom{00}4.03$   \\
 \hspace{2mm} \textbf{IPA}($r=2$)          &  $0.249$          & $199.94$  & $162.10$  &  $0.266$        & $244.42$ & $208.56$ \\
 \hspace{5mm} \textbf{PU}                  &  $0.048$          & $ \phantom{0}45.06$   & $ \phantom{0}36.79$   &  $0.008$        &  $ \phantom{00}7.02$  & $ \phantom{0}05.90$   \\
 \hspace{8mm} \textbf{\textit{Projection}}            &  $0.008$          & $ \phantom{00}5.98$    & $ \phantom{00}4.32$    &  $0.001$  & $\phantom{00}0.34$   & $\phantom{00}0.27$   \\
 \hspace{8mm} \textbf{\textit{FOD}}                 &  $0.032$          & $\phantom{0}27.35$   & $\phantom{0}21.69$   &  $0.006$        & $\phantom{00}5.45$   & $\phantom{00}4.46$   \\
 \hspace{8mm} \textbf{\textit{PreAggregation}}              &  $0.008$          & $\phantom{00}8.19$    & $\phantom{00}7.73$    &  $0.000$  & $\phantom{00}0.18$   & $\phantom{00}0.16$   \\
 \hspace{5mm} \textbf{Divider}             &  $0.047$          & $\phantom{0}25.64$   & $\phantom{0}23.04$   &  $0.023$        & $\phantom{0}25.94$  & $\phantom{0}25.20$  \\
 \hspace{5mm} \textbf{Register array}              &  $0.075$          & $\phantom{00}3.53$    & $\phantom{00}3.53$    &  $0.006$        & $\phantom{00}5.56$   & $\phantom{00}5.42$   \\
 \hspace{5mm} \textbf{Control unit}        &  $0.001$     & $\phantom{00}0.07$    & $\phantom{00}0.07$    &  $0.001$  & $\phantom{00}0.00$  & $\phantom{00}0.00$  \\
 \hspace{2mm} \textbf{\textit{PreAggregation}}($r=3$)       &   -               & -         & -         &  $0.005$        & $\phantom{00}4.39$   & $\phantom{00}3.22$   \\
 \hspace{2mm} \textbf{Divider}($r=3$)      &   -               & -         & -         &  $0.021$        & $\phantom{0}13.21$ & $\phantom{0}12.78$  \\
 \hspace{2mm} \textbf{Register array} ($r=3$)      &   -               & -         & -         &  $0.003$        & $\phantom{00}2.20$   & $\phantom{00}2.22$   \\
 \hspace{2mm} \textbf{Control unit}($r=3$) &   -               & -         & -         &  $0.000$  & $\phantom{00}0.01$   & $\phantom{00}0.01$  \\

\bottomrule
 
\end{tabular}}
{
\begin{tablenotes}[flushleft]
\item[a]Average area utilization for one iteration.
\end{tablenotes}
}

\end{threeparttable}
\end{adjustbox}

\end{table}
\section{Conclusion}
\label{sec:conclusion}
In this work, we described a pipelined and flexible architecture for {soft-decision} IPA decoding of RM codes. We used several algorithmic\footnote{We note that IPA decoding can be further simplified using recently published methods~\cite{JiaJie2021,Hashemipour2022}. Since our results, considered as a baseline, are already highly promising even without these simplifications, we consider the quantification of these additional  improvements as future work.} and architectural optimizations to reduce the hardware implementation complexity while maintaining an error-correcting performance that is very close to the original RPA decoding algorithm from which IPA is derived. 
{Our synthesis results with an STM $28$~nm technology demonstrate that the IPA decoder exhibits notable advantages in terms of area efficiency, with improvements of up to $10$ times, and latency reductions of up to $54\%$ when compared to the SCL decoder. These improvements are achieved while maintaining comparable error-correcting performance, highlighting the potential benefits of the IPA decoder for applications that require high reliability and low latency.  However, the post-synthesis simulation results showcases significantly higher power consumption compared to a state-of-the-art SCL decoder for polar codes. 
Therefore, the proposed flexible architecture of IPA enables a wide range of trade-offs between area and power consumption on one side, and latency and throughput on the other side.}

\label{sec:conc}

\bibliographystyle{IEEEtran}

\begin{thebibliography}{10}
\providecommand{\url}[1]{#1}
\csname url@samestyle\endcsname
\providecommand{\newblock}{\relax}
\providecommand{\bibinfo}[2]{#2}
\providecommand{\BIBentrySTDinterwordspacing}{\spaceskip=0pt\relax}
\providecommand{\BIBentryALTinterwordstretchfactor}{4}
\providecommand{\BIBentryALTinterwordspacing}{\spaceskip=\fontdimen2\font plus
\BIBentryALTinterwordstretchfactor\fontdimen3\font minus
  \fontdimen4\font\relax}
\providecommand{\BIBforeignlanguage}[2]{{%
\expandafter\ifx\csname l@#1\endcsname\relax
\typeout{** WARNING: IEEEtran.bst: No hyphenation pattern has been}%
\typeout{** loaded for the language `#1'. Using the pattern for}%
\typeout{** the default language instead.}%
\else
\language=\csname l@#1\endcsname
\fi
#2}}
\providecommand{\BIBdecl}{\relax}
\BIBdecl

\bibitem{Mahmood2020}
N.~H. {Mahmood \emph{et al.}}, ``White paper on critical and massive machine
  type communication towards {6G},'' \emph{arXiv 2004.14146}, 2020.

\bibitem{Durisi2016}
G.~Durisi, T.~Koch, and P.~Popovski, ``Toward massive, ultrareliable, and
  low-latency wireless communication with short packets,'' \emph{Proceedings of
  the IEEE}, pp. 1711--1726, Sep 2016.

\bibitem{Chen2018}
H.~{Chen \emph{et al.}}, ``Ultra-reliable low latency cellular networks: Use
  cases, challenges and approaches,'' \emph{{IEEE} Communications Magazine},
  vol.~56, no.~12, pp. 119--125, 2018.

\bibitem{gallager1963low}
R.~Gallager, ``Low density parity check codes,'' \emph{Cambridge}, vol.~1, pp.
  1--73, 1963.

\bibitem{berrou1996near}
C.~Berrou and A.~Glavieux, ``Near optimum error correcting coding and decoding:
  Turbo-codes,'' \emph{{IEEE} Transactions on communications}, vol.~44, no.~10,
  pp. 1261--1271, 1996.

\bibitem{Arikan2009}
E.~Arikan, ``Channel polarization: A method for constructing capacity-achieving
  codes for symmetric binary-input memoryless channels,'' \emph{{IEEE}
  Transactions on Information Theory}, vol.~55, no.~7, pp. 3051--3073, jul
  2009.

\bibitem{Coskun2019}
M.~C. {Co{\c{s}}kun \emph{et al.}}, ``Efficient error-correcting codes in the
  short blocklength regime,'' \emph{Physical Communication}, vol.~34, pp.
  66--79, Jun. 2019.

\bibitem{Tonnellier2021}
T.~Tonnellier, M.~Hashemipour-Nazari, N.~Doan, W.~J. Gross, and
  A.~Balatsoukas-Stimming, ``Towards practical near-maximum-likelihood decoding
  of error-correcting codes: An overview,'' in \emph{{IEEE} International
  Conference on Acoustics, Speech and Signal Processing ({ICASSP})}, Jun. 2021.

\bibitem{Reed1954}
I.~Reed, ``A class of multiple-error-correcting codes and the decoding
  scheme,'' \emph{Transactions of the {IRE} Professional Group on Information
  Theory}, vol.~4, no.~4, pp. 38--49, Sep 1954.

\bibitem{Muller1954}
D.~E. Muller, ``Application of boolean algebra to switching circuit design and
  to error detection,'' \emph{Transactions of the IRE Professional Group on
  Electronic Computers}, vol. {EC}-3, no.~3, pp. 6--12, Sep 1954.

\bibitem{Dumer2004}
I.~Dumer, ``Recursive decoding and its performance for low-rate
  {Reed{\textendash}Muller} codes,'' \emph{{IEEE} Transactions on Information
  Theory}, vol.~50, no.~5, pp. 811--823, May 2004.

\bibitem{Sakkour2005}
B.~Sakkour, ``Decoding of second order {Reed{\textendash}Muller} codes with a
  large number of errors,'' in \emph{{IEEE} Information Theory Workshop}, Aug
  2005.

\bibitem{Dumer2006a}
I.~Dumer, ``Soft-decision decoding of {Reed{\textendash}Muller} codes: {A}
  simplified algorithm,'' \emph{{IEEE} Transactions on Information Theory},
  vol.~52, no.~3, pp. 954--963, Mar 2006.

\bibitem{Dumer2006}
I.~Dumer and K.~Shabunov, ``Soft-decision decoding of {Reed{\textendash}Muller}
  codes: {Recursive lists},'' \emph{{IEEE} Transactions on Information Theory},
  vol.~52, no.~3, pp. 1260--1266, Mar 2006.

\bibitem{saptharishi2017}
R.~Saptharishi, A.~Shpilka, and B.~L. Volk, ``Efficiently decoding
  {Reed-Muller} codes from random errors,'' in \emph{{IEEE} Transactions on
  Information Theory}, vol.~63, 2017, pp. 1954--1960.

\bibitem{Costello2007}
D.~J. Costello and G.~D. Forney, ``Channel coding: The road to channel
  capacity,'' \emph{Proceedings of the IEEE}, vol.~95, no.~6, pp. 1150--1177,
  Jun 2007.

\bibitem{Abbe2015}
E.~Abbe, A.~Shpilka, and A.~Wigderson, ``{Reed{\textendash}Muller} codes for
  random erasures and errors,'' in \emph{Annual {ACM} Symposium on Theory of
  Computing}, Jun 2015.

\bibitem{Kudekar2017}
S.~Kudekar, S.~Kumar, M.~Mondelli, H.~D. Pfister, E.~Sasoglu, and R.~L.
  Urbanke, ``{Reed{\textendash}Muller} codes achieve capacity on erasure
  channels,'' \emph{{IEEE} Transactions on Information Theory}, vol.~63, no.~7,
  pp. 4298--4316, Jul 2017.

\bibitem{Abbe2019}
E.~Abbe and M.~Ye, ``{Reed{\textendash}Muller} codes polarize,'' \emph{IEEE
  Transactions on Information Theory}, vol.~66, no.~12, pp. 7311--7332, 2020.

\bibitem{Sberlo2020}
O.~Sberlo and A.~Shpilka, ``On the performance of {Reed{\textendash}Muller}
  codes with respect to random errors and erasures,'' in \emph{Annual ACM-SIAM
  Symposium on Discrete Algorithms}, 2020, pp. 1357–--1376.

\bibitem{Arkan2009}
E.~Arıkan, H.~Kim, G.~Markarian, {\"U}.~{\"O}zg{\"u}r, and E.~Poyraz,
  ``Performance of short polar codes under {ML} decoding,'' in
  \emph{ICT-MobileSummit Conference}, Sep. 2009, pp. 10--12.

\bibitem{Mondelli2014}
M.~Mondelli, S.~H. Hassani, and R.~L. Urbanke, ``From polar to
  {Reed{\textendash}Muller} codes: A technique to improve the finite-length
  performance,'' \emph{IEEE Transactions on Communications}, vol.~62, no.~9,
  pp. 3084--3091, Sep 2014.

\bibitem{Ye2020}
M.~Ye and E.~Abbe, ``Recursive projection-aggregation decoding of
  {Reed{\textendash}Muller} codes,'' \emph{{IEEE} Transactions on Information
  Theory}, vol.~66, no.~8, pp. 4948--4965, Aug 2020.

\bibitem{Lian2020}
M.~Lian, C.~Hager, and H.~D. Pfister, ``Decoding {Reed{\textendash}Muller}
  codes using redundant code constraints,'' in \emph{{IEEE} International
  Symposium on Information Theory ({ISIT})}, Jun 2020.

\bibitem{Huang2022}
Q.~Huang and B.~Zhang, ``Pruned collapsed projection-aggregation decoding of
  {Reed-Muller} codes,'' \emph{arXiv:2105.11878}, May 2021.

\bibitem{li2022optimization}
J.~Li and W.~J. Gross, ``Optimization and simplification of {PCPA} decoder for
  {Reed-Muller} codes,'' \emph{IEEE Communications Letters}, vol.~26, no.~6,
  pp. 1206--1210, Jun 2022.

\bibitem{Fathollahi2021}
D.~Fathollahi, N.~Farsad, S.~A. Hashemi, and M.~Mondelli, ``Sparse
  multi{\textendash}decoder recursive projection aggregation for
  {Reed{\textendash}Muller} codes,'' in \emph{{IEEE} International Symposium on
  Information Theory}, Jul 2021.

\bibitem{JiaJie2021}
J.~Li, S.~M. Abbas, T.~Tonnellier, and W.~J. Gross, ``Reduced complexity {RPA}
  decoder for {Reed{\textendash}Muller} codes,'' in \emph{{IEEE} International
  Symposium on Topics in Coding}, Sep 2021.

\bibitem{Hashemipour-Nazari2021}
M.~Hashemipour-Nazari, K.~Goossens, and A.~Balatsoukas-Stimming, ``Hardware
  implementation of iterative projection-aggregation decoding of
  {Reed{\textendash}Muller} codes,'' in \emph{{IEEE} International Conference
  on Acoustics, Speech and Signal Processing ({ICASSP})}, Jun 2021.

\bibitem{Tal2015}
I.~Tal and A.~Vardy, ``List decoding of polar codes,'' \emph{IEEE Transactions
  on Information Theory}, vol.~61, no.~5, pp. 2213--2226, 2015.

\bibitem{Balatsoukas-Stimming2015}
A.~Balatsoukas-Stimming, M.~{Bastani Parizi}, and A.~Burg, ``{LLR}-based
  successive cancellation list decoding of polar codes,'' \emph{{IEEE}
  Transactions on Signal Processing}, vol.~63, no.~19, pp. 5165--5179, Oct
  2015.

\bibitem{Ren2022}
Y.~Ren, A.~T. Kristensen, Y.~Shen, A.~Balatsoukas-Stimming, C.~Zhang, and
  A.~Burg, ``A sequence repetition node-based successive cancellation list
  decoder for {5G} polar codes: {Algorithm} and implementation,'' 
\emph{{IEEE} Transactions on Signal Processing}, vol.~70, pp. 5592-5607, 2022.

\bibitem{MacWilliams}
F.~MacWilliams and N.~Sloane, \emph{The Theory of Error-Correcting
  Codes}.\hskip 1em plus 0.5em minus 0.4em\relax Amsterdam, The Netherlands:
  Elsevier, 1977.

\bibitem{Abbe2021}
E.~Abbe, A.~Shpilka, and M.~Ye, ``Reed{\textendash}{M}uller codes: Theory and
  algorithms,'' \emph{{IEEE} Transactions on Information Theory}, vol.~67,
  no.~6, pp. 3251--3277, Jun 2021.

\bibitem{Green66}
R.~Green, ``A serial orthogonal decoder,'' \emph{Jet Propulsion Laboratory
  (JPL) Space Programs Summary}, vol.~37, pp. 247--253, 1966.

\bibitem{Beext86}
Y.~Be'ery and J.~Snyders, ``Optimal soft decision block decoders based on fast
  {H}adamard transform,'' \emph{{IEEE} Transactions on Information Theory},
  vol.~32, no.~3, pp. 355--364, May 1986.

\bibitem{Fossorier1999}
M.~Fossorier, M.~Mihaljevic, and H.~Imai, ``Reduced complexity iterative
  decoding of low-density parity check codes based on belief propagation,''
  \emph{IEEE Transactions on Communications}, vol.~47, no.~5, pp. 673--680, May
  1999.

\bibitem{Zheng21}
H.~{Zheng \emph{et al.}}, ``Threshold-based fast successive-cancellation
  decoding of polar codes,'' \emph{IEEE Transactions on Communications},
  vol.~69, no.~6, pp. 3541--3555, 2021.

\bibitem{matlab5gnr}
\BIBentryALTinterwordspacing
{MathWorks}. (2021) {5G} new radio polar coding. Accessed: 2022-09-30.
  [Online]. Available:
  \url{https://nl.mathworks.com/help/5g/gs/polar-coding.html}
\BIBentrySTDinterwordspacing

\bibitem{Hashemipour2022}
M.~Hashemipour-Nazari, K.~Goossens, and A.~Balatsoukas-Stimming, ``Multi-factor
  pruning for recursive projection-aggregation decoding of {RM} codes,'' in
  \emph{IEEE Workshop on Signal Processing Systems (SiPS)}, Nov 2022.

\end{thebibliography}

\end{document}